\documentclass[usenatbib]{mnras}
\usepackage{aas_macros}
\usepackage{amsmath}
\usepackage{amssymb}
\usepackage{color}
\usepackage{epsfig}
\usepackage{float}
\usepackage{graphicx}
\usepackage{latexsym}
\usepackage{morefloats}
\usepackage{natbib}
\usepackage{times}

\newcommand{\plotone}[1]{\resizebox{0.95\hsize}{!}{\includegraphics{#1}}}

\newcommand{\Msun}{{~\rm M_\odot}}

\newcommand{\kpc}{~\rm kpc}
\newcommand{\Mpc}{~\rm Mpc}

\newcommand{\uvec}[1]{\boldsymbol{\mathit{\hat{#1}}}}
\renewcommand{\vec}[1]{\boldsymbol{\mathit{#1}}}

\newcommand{\td}{\textendash}

\def\gsim { \lower .75ex \hbox{$\sim$} \llap{\raise .27ex \hbox{$>$}}}
\def\lsim { \lower .75ex \hbox{$\sim$} \llap{\raise .27ex \hbox{$<$}}}
\newcommand{\eagle}{\textsc{eagle}}
\newcommand{\simRef}{Ref-L{\small 0100}N{\small 1504}}
\newcommand{\lcdm}{$\Lambda$CDM}

\newcommand{\reffig}[1]{Fig. \ref{#1}}

\title[Multiscale galactic alignments]
{Alignments between galaxies, satellite systems and haloes}
\author[Shi Shao et al.]
{\parbox{\textwidth}{
        Shi Shao$^{1,2}$ \thanks{E-mail : shaoshi@bao.ac.cn},
        Marius Cautun$^{2}$\thanks{E-mail : m.c.cautun@durham.ac.uk},
        Carlos S.~Frenk$^{2}$,
        Liang Gao$^{1,2}$,
        Robert A. Crain$^{3}$,
        Matthieu Schaller$^{2}$,
        Joop Schaye$^{4}$ and
        Tom Theuns$^{2}$
        \vspace{.20cm}} \\
$^1$National Astronomical Observatories, The Partner Group of Max Planck Institute for Astrophysics, Chinese Academy of Sciences, Beijing, 100012, China\\
$^2$Institute for Computational Cosmology, Department of Physics, Durham University, South Road Durham DH1 3LE, UK \\
$^3$Astrophysics Research Institute, Liverpool John Moores University, IC2, 146 Brownlow Hill, Liverpool, L3 5RF, UK\\
$^4$Leiden Observatory, Leiden University, PO Box 9513, NL-2300 RA Leiden, the Netherlands
}

\begin{document}
\maketitle
\begin{abstract}
    The spatial distribution of the satellite populations of the Milky
    Way and Andromeda are puzzling in that they are nearly perpendicular
    to the discs of their central galaxies. To understand the origin of
    such configurations we study the alignment of the central galaxy,
    satellite system and dark matter halo in the largest of the `Evolution
    and Assembly of GaLaxies and their Environments' (\eagle{})
    simulation. We find that centrals and their satellite systems tend to
    be well aligned with their haloes, with a median misalignment angle of
    $33^\circ$ in both cases. While the centrals are better aligned with
    the inner $10\kpc$ halo, the satellite systems are better aligned with
    the entire halo indicating that satellites preferentially trace the
    outer halo. The central--satellite alignment is weak (median
    misalignment angle of $52^\circ$) and we find that around $20\%$ of
    systems have a misalignment angle larger than $78^\circ$, which is the
    value for the Milky Way. The central--satellite alignment is a
    consequence of the tendency of both components to align with the dark
    matter halo. As a consequence, when the central is parallel to the
    satellite system, it also tends to be parallel to the halo. In
    contrast, if the central is perpendicular to the satellite system, as
    in the case of the Milky Way and Andromeda, then the central--halo
    alignment is much weaker. Dispersion-dominated (spheroidal) centrals
    have a stronger alignment with both their halo and their satellites
    than rotation-dominated (disc) centrals. We also found that the halo,
    the central galaxy and the satellite system tend to be aligned with
    the surrounding large-scale distribution of matter, with the halo
    being the better aligned of the three.
\end{abstract}

\begin{keywords}
methods: numerical - galaxies: haloes - galaxies: kinematics and dynamics
\end{keywords}

\section{Introduction} \label{sect:intro}
The distribution of galactic satellites is highly inhomogeneous and
anisotropic, as can be easily recognized from observations of the
Local Group (LG). Most of the Milky Way (MW) satellites define a tight
plane \citep{Kunkel_76, Lynden-Bell_76, Lynden-Bell_82, Kroupa_05}
that shows some degree of coherent rotation \citep{Metz_08,
Pawlowski_13}. Even more puzzling is the orientation of this satellite
plane which is almost perpendicular to the MW disc. The satellites of
Andromeda (M31) are distributed mostly along two planar, nearly
parallel structures that are offset from each other
\citep{Conn_13,Ibata_13,Shaya2013}. Such planes of satellites are
common outside the LG too \citep[e.g. the Centaurus A
Group,][]{Tully2015} with their characteristic signature detected in
large stacked samples of external galaxies \citep{Cautun2015b}.

Within the standard $\Lambda$ cold dark matter (\lcdm{}) model, the
anisotropic distribution of satellites is a manifestation of the
preferential direction of accretion on to haloes
\citep[e.g.][]{Aubert_04,Knebe_04,Libeskind_05,Zentner_05b,
Li_08,Deason2011,Wang_14,Shi_15}. The flattened distributions of
satellites can arise from the infall of satellites along the spine of
filaments \citep{Libeskind_05,Buck_15} and that can also lead to a
significant population of corotating satellites
\citep{Libeskind_09,Lovell2011,Cautun2015b}. Despite \lcdm{}
predicting the existence of satellite planes, initial studies
emphasized a perceived discrepancy with observations, with the MW and
M31 satellite planes claimed to be thinner and to show a larger degree
of coherent rotation than their \lcdm{} counterparts
\citep[e.g.][]{Pawlowski2012b,Ibata2014b}. However, \citet{Cautun_15}
recently pointed out that this conclusion was based on a
misinterpretation of the diversity of satellite planes \citep[see
also][]{Buck_16}, since the characteristics of each plane
(e.g. thickness, radial extent) vary strongly from halo to halo. In
fact, the very diversity of satellite planes is a manifestation of the
varied formation and evolution history of the host halo
\citep[e.g. see][]{Buck_15,Smith2016}.

In this paper, we investigate the information encoded in the
preferential direction of the spatial distributions of satellites,
focusing on the alignment of satellite planes with the direction of
the central galaxies and host haloes. Within the \lcdm{} cosmological
model, dark matter (DM), gas and satellites are accreted
preferentially along filaments suggesting that these various
subsystems should be aligned to some extent
\citep{Libeskind_05,Libeskind2011,Libeskind2014}. Most studies have
focused on two aspects of these correlations. First, hydrodynamical
simulations show that the central galaxy has a typical misalignment
angle of $\approx 30^\circ$ with the DM halo, with an even stronger
alignment for spheroids \citep[e.g.][]{Bett_10,Deason2011,
Sales_12,Tenneti_14,Velliscig_15a}. Secondly, both observations and
simulations show that individual satellites are preferentially aligned
along the major axis of the central galaxy, with the strongest
alignment occurring between red satellites and red centrals
\citep[e.g.][]{Brainerd_05,Yang_06,Agustsson2010,Nierenberg2012,
Dong_14,Tenneti_14,Velliscig_15b}. However, the alignment of the whole
distribution of satellites with the central galaxy and with the DM
halo, which is the focus of this study, has been largely overlooked
\citep[although see][]{Libeskind2007,Libeskind_09,Deason2011},
despite its importance for interpreting the LG observations. The
satellite systems of both the MW and M31 are roughly perpendicular to
the disc of their respective centrals and are thus difficult to
reconcile with the expectation of the filamentary accretion
hypothesis. To address this puzzle we will determine the prevalence of
such perpendicular configurations and study their implications.

Our study makes use of the hydrodynamical simulations run as part
of the `Evolution and Assembly of GaLaxies and their Environments'
(\eagle{}) project (\citealt{Crain_15, Schaye_15}). \eagle{}
implements the main physical processes that determine the formation
and evolution of galaxies, incorporating the baryonic processes that
affect the galaxy and halo shapes as well as the orbits of satellite
galaxies. This simulation was used by \citet{Velliscig_15a} to study
the alignments of the distributions of stars, hot gas and DM. They
found that, while galaxies are well aligned with the local
distribution of DM, they can have large misalignments with the entire
halo. In a separate study, \citet{Velliscig_15b} used the same
simulation to measure that the strength of the galaxy-galaxy alignment
is a strongly decreasing function of the distance between the two
objects \citep[see also][]{Welker2015}. By contrast, our study focuses
on the alignment of satellite systems and on its interplay with the
central galaxy, the host halo and the surrounding distribution of
matter.

The paper is organized as follows. Section~\ref{sect:simul} reviews
the \eagle{} simulation and describes our sample selection;
Section~\ref{sect:result} presents our main results;
Section~\ref{sect:disc} discusses the implications of our findings; we
conclude with a short summary in Section~\ref{sect:conclusion}.

\section{Simulation and methods} 
\label{sect:simul}
We make use of the main cosmological hydrodynamical simulation
(labelled \simRef{}) performed as part of the \eagle{} project
\citep{Schaye_15, Crain_15}. \eagle{} assumes a \textit{Planck} cosmology
\citep{Planck2014} with cosmological parameters: $\Omega_{\rm m}=0.307,
\Omega_{\rm b}=0.04825,\Omega_\Lambda=0.693,h=0.6777,\sigma_8=0.8288$ and
$n_{\rm s}=0.9611$. The simulation is of a periodic cube of
$100\Mpc{}$ side length and follows the evolution $1504^3$
DM and an initially equal number of baryonic particles. The DM
particles have a mass of $9.7\times 10^6 \Msun$, while the gas
particles have an initial mass of $1.8\times 10^6 \Msun$.

The simulation was performed using a version of the \textsc{gadget}
code \citep{Springel_05} which has been modified to include
state-of-the-art smoothed particle hydrodynamics methods \citep[Dalla
Vecchia, in preparation]{Hopkins2013,Schaller2015a}. The baryonic physics
implementation includes element-by-element cooling using the
\citet{Wiersma2009a} prescription in the presence of a
\citet{Haardt2001} ultraviolet (UV) and X-ray background, stochastic star formation
with a metallicity dependent threshold \citep{Schaye2004} and a star
formation rate that depends explicitly on pressure \citep{Schaye2008},
thermal energy feedback associated with star formation
\citep{DallaVecchia2012}, and the injection of hydrogen, helium and
metals into the interstellar medium from supernovae and stellar mass
loss \citep{Wiersma2009b}. Star particles are treated as single
stellar populations with a \citet{Chabrier2003} initial mass
function. Supermassive black holes grow through mergers and accretion
of low angular momentum material
\citep{Springel_05a,Rosas-Guevara2015,Schaye_15} and the resulting
active galactic nuclei (AGN) inject thermal energy into the surrounding gas
\citep{Booth2009,DallaVecchia2012}. These subgrid models were
calibrated to reproduce the present day stellar mass function and
galaxy sizes, as well as the relation between galaxy stellar masses
and supermassive black hole masses \citep{Crain_15, Schaye_15}. See
\citet{Schaye_15} for a more detailed description of the baryonic
processes implemented in \eagle{}.

Haloes are identified using the friends-of-friends (FOF) algorithm
with a linking length of $0.2$ times the mean particle separation
\citep{Davis1985}. Gravitationally bound substructures are identified
using the \textsc{subfind} code \citep{Springel_01,Dolag2009} applied
to the full matter distribution (DM, gas and stars) associated with each
FOF halo. The subhalo that contains the particle with the lowest
gravitational energy is classified as the main halo and its stellar
distribution as the central galaxy. The main haloes are characterized
by the mass, $M_{200}$, and radius, $R_{200}$, that define an enclosed
spherical overdensity of $200$ times the critical density. The
remaining subhaloes are classified as satellite galaxies. The position
of each galaxy, for both centrals and satellites, is given by the
particle that has the lowest gravitational potential energy.

\subsection{Sample selection}
\begin{figure}                 
  \plotone{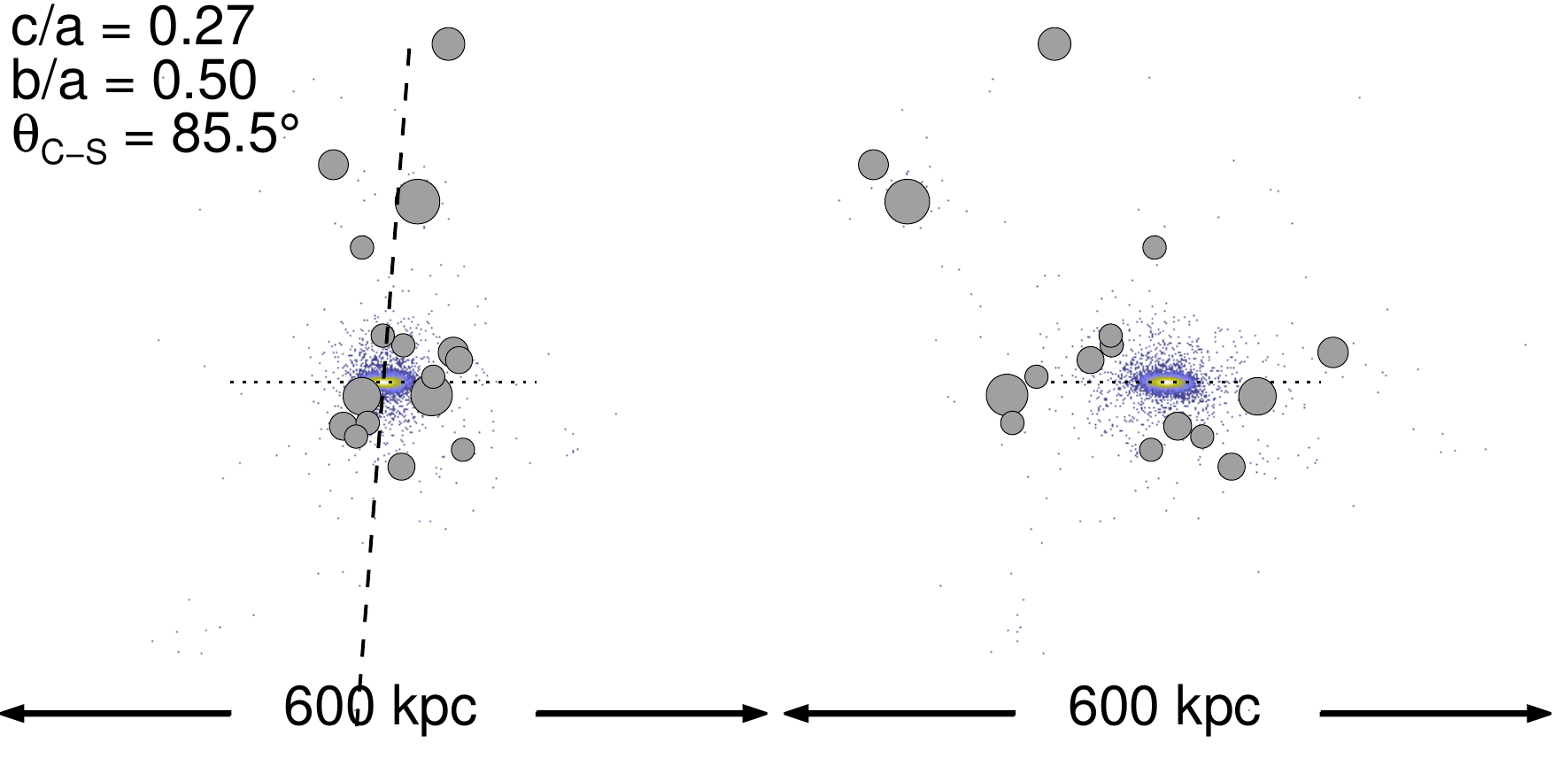}
  \plotone{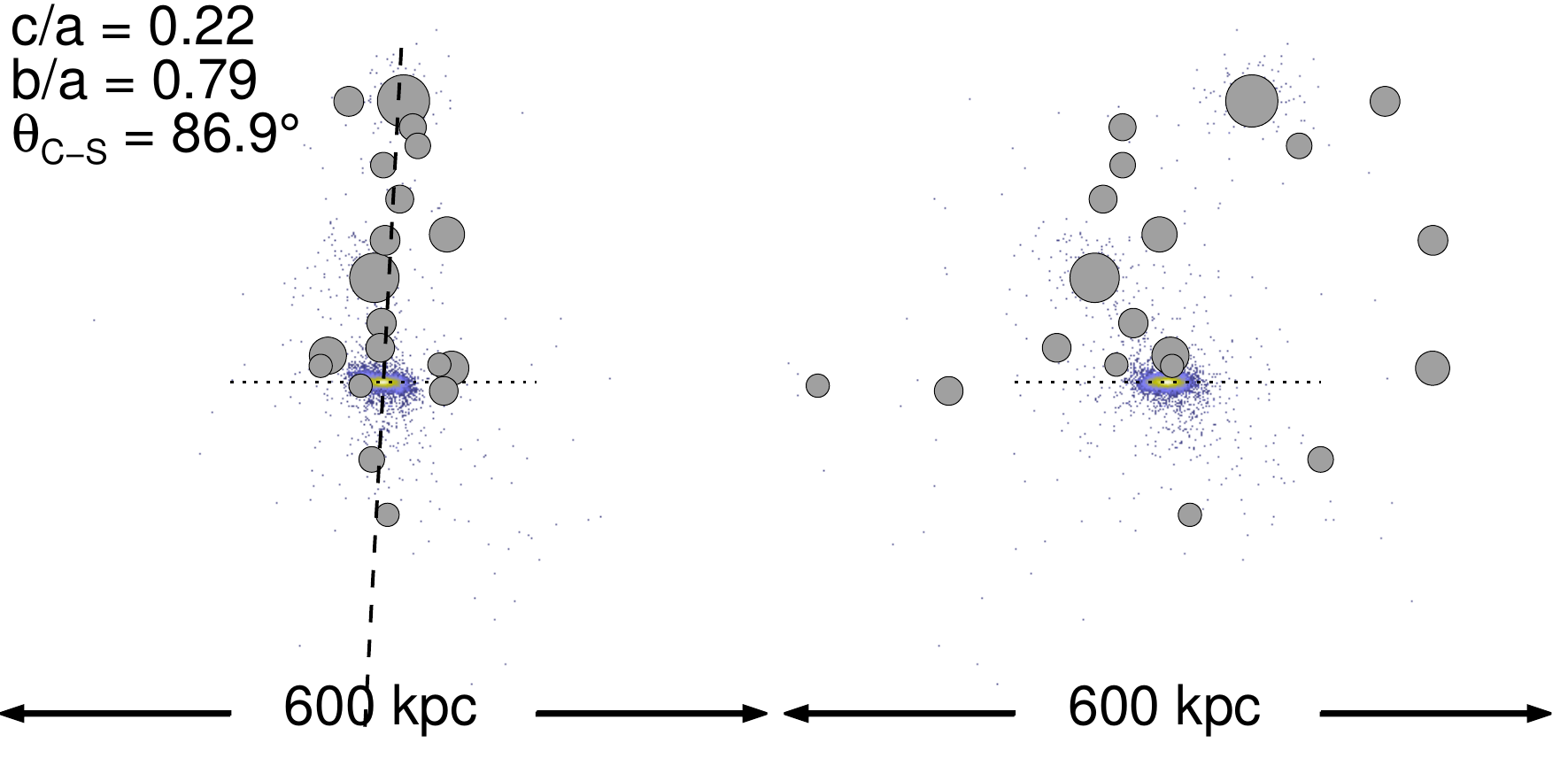}
  \plotone{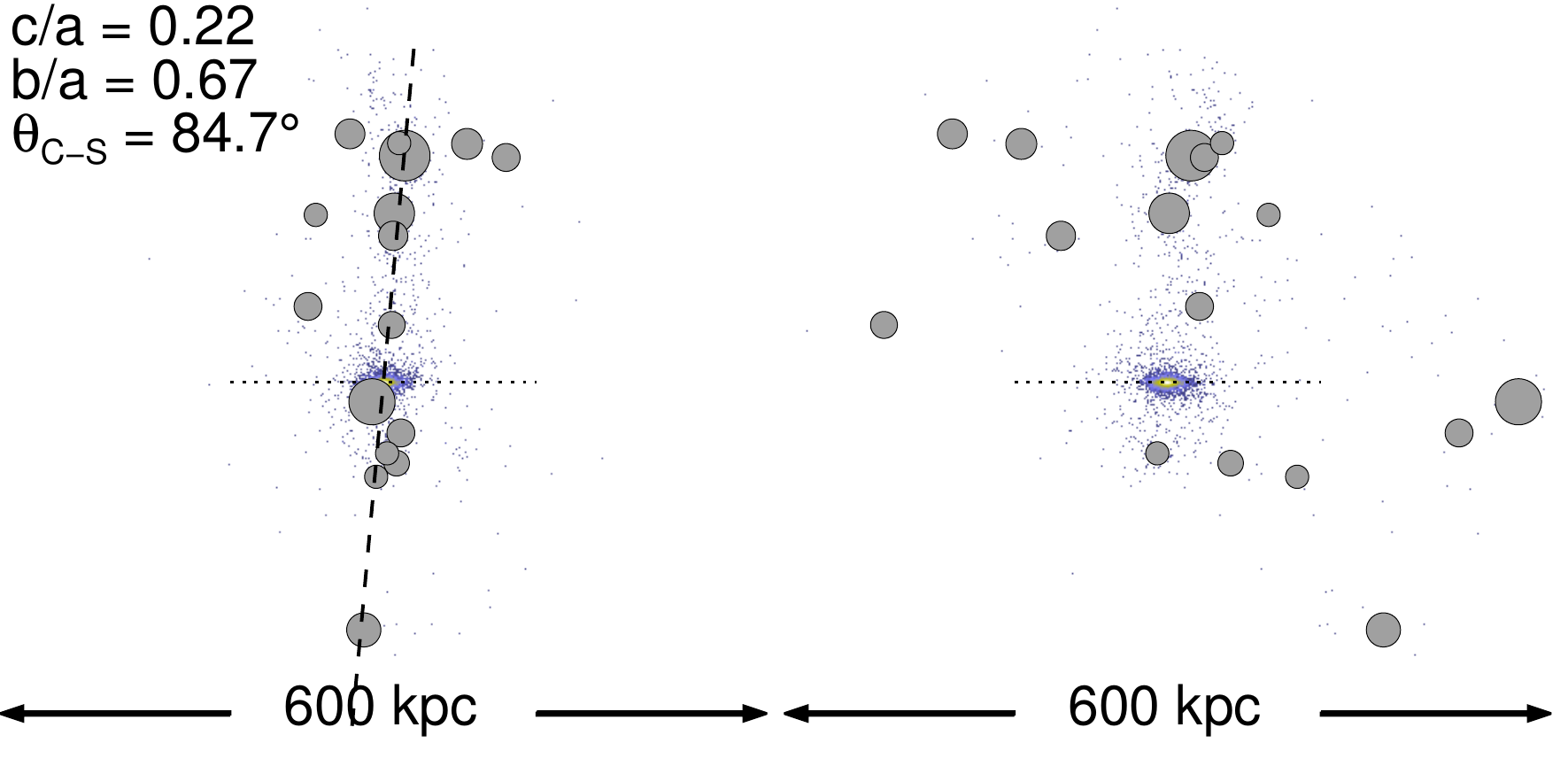}
  \plotone{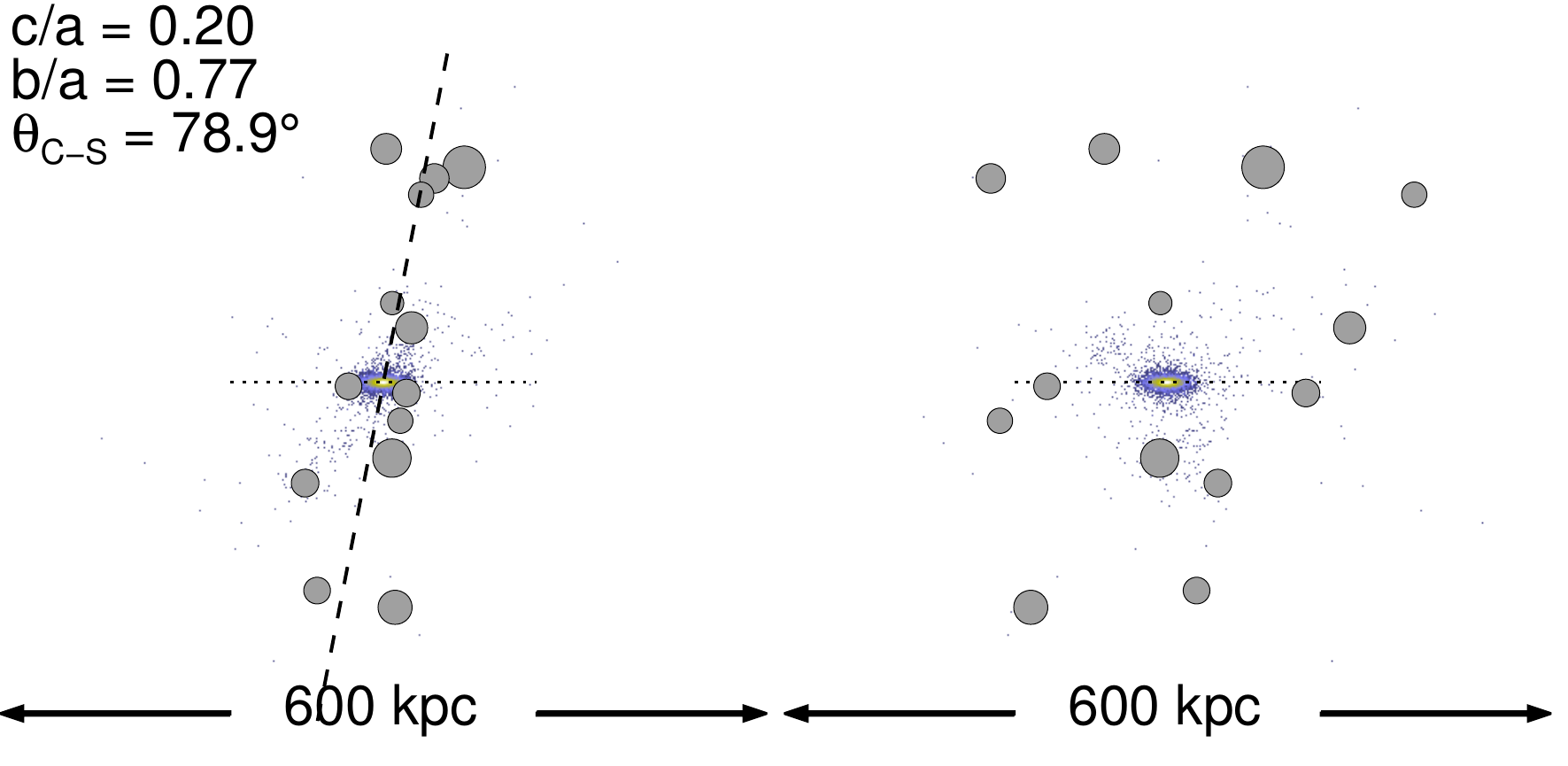}
  \plotone{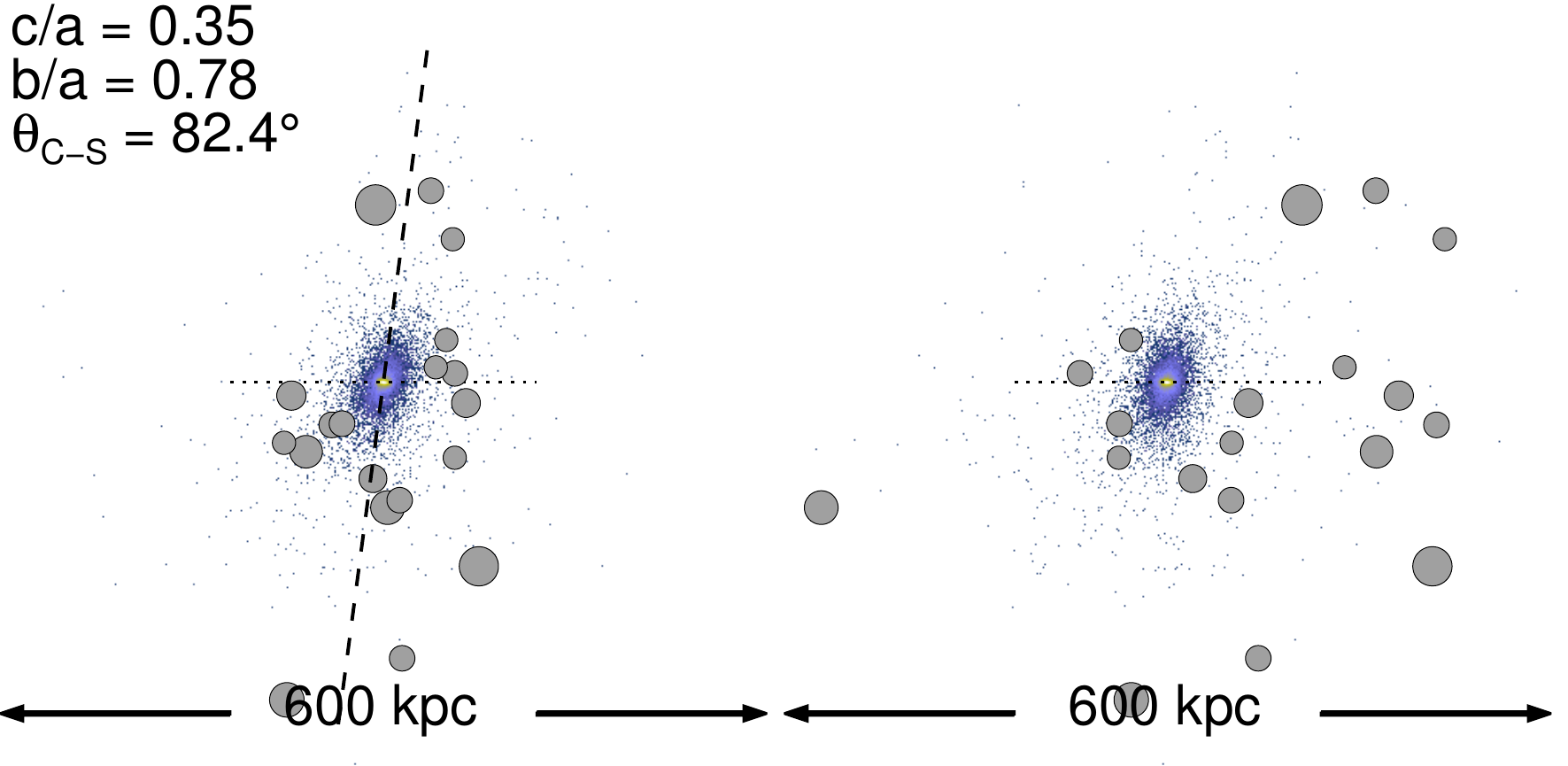}
  \caption{Examples of galactic systems that have planar satellite
    distributions that are almost perpendicular to the central
    galaxy. Each row shows a different system, with the two columns
    showing orthogonal projections. The two projections are edge-on views
    of the central galaxy, with the satellite plane being shown edge-on
    and face-on in the left- and right-hand columns, respectively. The blue hue
    shows the distribution of stars which is dominated by the central
    galaxy. The luminous satellites are shown as grey circles with sizes
    varying according to their stellar mass. The dotted line shows the
    best-fitting plane of the central galaxy while the dashed line in the
    left-hand column shows the best-fitting plane of the satellite system. The
    top left-hand text insert gives the shape of the satellite system and its
    angle with respect to the central galaxy.
  }
  \label{fig:P1}
\end{figure}

To identify systems similar to the MW and M31, we start by selecting
the 3209 haloes with mass $M_{200} \in [0.3, 3] \times
10^{12}\Msun$. The wide mass range is motivated by the large
uncertainties in the mass of the MW \citep[e.g.][]{Fardal_13,
Cautun_14a, Piffl_14, Wang_15, Han_16} and the need to have a large
sample of such systems. We require that any such halo be isolated and
not overlap with more massive companions. Thus, we exclude all central
galaxies that have a neighbour within $600\kpc$ that has a stellar
mass larger than half their mass. We also restrict our selection to
haloes that, like the MW, have at least 11 luminous satellites within
a distance of $300 \kpc$ from the central galaxy. A luminous satellite
consists of a DM subhalo with at least one star particle.  We obtain
1080 host haloes that satisfy all three selection criteria.  The
sample has a median halo mass, $M_{200} \sim 1.2 \times 10^{12}\Msun$,
and a median number of $15$ luminous satellites per halo. The typical
total mass of a luminous satellite is $M_{\rm tot} \sim 1
\times10^{9} \Msun$, which corresponds to $\sim 100$ DM particles (see
Appendix~\ref{sect:appendix_sample} for the halo and satellite mass
functions). \reffig{fig:P1} illustrates the distribution of stars and
satellites in five haloes found in our sample. These systems, which we
will discuss in detail in Section~\ref{sect:result}, were selected to
have satellite system that are almost perpendicular to the central
galaxy, similar to the configuration observed around the MW.

\subsection{Shape definition}
\label{subsec:shape_definition}

We compute the shape of the various galactic subsystems (e.g. halo,
central galaxy, satellite population) using the moment of inertia
tensor\footnote{Strictly speaking, $I_{ij}$, is not the moment of
inertia tensor \citep[see e.g.][]{Bett2007}, but we follow the common
practice in this subject and adopt this nomenclature.},
\begin{equation}
  I_{ij} \equiv \sum_{k=1}^{N} m_{k} x_{k,i} x_{k,j} \;,
  \label{eq:inertia_tensor}
\end{equation}
where $N$ is the number of particles that belong to the structure of
interest, $x_{k,i}$ denotes the $i$-th component ($i=1,2,3$) of the
position vector of particle $k$ with respect to the halo centre, and
$m_k$ denotes the mass of that particle. In the case of the halo, the
sum is over all DM particles within $R_{200}$, while for the central
galaxy the sum is over all the star particles within $10\kpc$ from the
centre (which is approximately twice the average value of the half
stellar mass radius in our sample). For the satellite system, the sum
is over all the luminous satellites within $300\kpc$ of the centre,
with each satellite being assigned a constant and equal mass,
$m_k=1$. We weight all the satellites equally to compare more closely
to observations, where the satellite masses are highly uncertain, and
to use the same approach as previous works which have studied planes
of satellite galaxies
\citep[e.g.][]{Libeskind_05,Libeskind2007,Pawlowski2013b,Wang_13}.

The shape and the orientation are determined by the eigenvalues,
$\lambda_i$ ($\lambda_{1}\geqslant\lambda_{2}\geqslant\lambda_{3}$),
and the eigenvectors, $\uvec{e}_i$, of the inertia tensor. The major,
intermediate and minor axes of the corresponding ellipsoid are given
by $a=\sqrt{\lambda_1}$, $b=\sqrt{\lambda_2}$ and
$c=\sqrt{\lambda_3}$, respectively. The computation of the inertia
tensor using a spherical region biases the shape towards higher
sphericity, but this has little effect on the orientation of the
principal axes, which is the focus of our study
\citep{Frenk1988,Bailin2005}.

\subsection{The misalignment angle}
We are interested in the degree of alignment between the galactic
subsystems, which we will quantify in terms of a misalignment angle,
$\theta$. For example, the misalignment angle between the central
galaxy and its parent halo is defined as,
\begin{equation}
  \theta_{\rm C \td H} = \arccos( \; |\uvec{e}_3^{\rm C}\cdot\uvec{e}_3^{\rm H}| \;) \;,
  \label{eq:misalignment_angle}
\end{equation}
where $\uvec{e}_3^{\rm C}$ and $\uvec{e}_3^{\rm H}$ are the minor axes
of the central galaxy and the halo, respectively. Note that we take
the absolute value of the dot product because the eigenvectors
determine only an orientation and do not have a direction assigned to
them. The misalignment angles between the satellite plane and the
halo, $\theta_{\rm S \td H}$, and between the central galaxy and the
satellite plane, $\theta_{\rm C \td S}$, are computed similarly.

We focus our analysis on the misalignment angle between the minor axes
because a large fraction of central galaxies are discs and hence have
$a \approx b$ which makes it difficult to identify robustly the major
and intermediate axes (see \reffig{fig:dis-CH}). In contrast,
$c \leqslant b$ for all systems, independently of whether we measure the
shape of the halo, central galaxy or satellite system. While not
discussed, we have also studied the alignment between the major axes
of the various components and found it to be weaker than the alignment
of the minor axes, while the intermediate axes show a very weak
alignment, if any at all.

\subsection{Disc and spheroid galaxy samples}
\label{sect:kappa}
We split the centrals into disc and spheroidal galaxies, following the
procedure of \citet[][see also \citealt{Sales_12}]{Scannapieco_09} and
divide our sample according to the degree of ordered rotation. We
define the parameter, $\kappa_{\rm rot}$, as the fraction of kinetic
energy, $K$, invested in ordered rotation, i.e.
\begin{equation}
  \kappa_{\rm rot} \equiv \frac{K_{\rm rot}}{K} = \frac{
  \sum_j \frac{1}{2}m_j [(\uvec{L} \times \uvec{r}_j) \cdot \vec{v}_j]^2 } {
  \sum_j \frac{1}{2}m_j \vec{v}_j^2 } \;,
  \label{eq:kappa_rot}
\end{equation}
where $\vec{v}_j$, $\uvec{r}_j$ and $m_j$ are the velocity, unit
position vector and mass of the $j$th star particle in the centre of
mass reference frame and $\uvec{L}$ is the direction of the total
angular momentum of the stellar component of central galaxy. For
perfect circular motion $\kappa_{\rm rot} = 1$, while for non-rotating
systems, $\kappa_{\rm rot} \ll 1$. In practice, we classify the
galaxies with $\kappa_{\rm rot} \geqslant 0.6$ and $\kappa_{\rm rot}
\leqslant 0.45$ as discs and spheroids, respectively. This
classification results in roughly a third disc galaxies, another third
spheroids and the remaining third an intermediate population.

Note that our disc versus spheroid kinematic decomposition differs
from the customary photometry-based method used in observational
studies, with the two showing a moderate correlation with considerable
scatter \citep{Abadi_03,Scannapieco_10}. Applying the latter method to
simulations requires the creation of realistic galaxy images which
introduces an additional layer of complexity. We therefore restrict
our analysis to galaxies with high and low values of $\kappa_{\rm
rot}$, which correspond to the most disc- and spheroid-like objects.

\section{Results}
\label{sect:result}
In this section we determine the alignment between the three galactic
subsystems: the central galaxy, the DM halo and the satellite
system. Our analysis is based on haloes with masses similar to that of
the MW halo for which \eagle{} has just the right volume to include a
large number of such objects while having enough resolution to detect
their bright satellite populations. We will also characterize the
alignment of these galactic subsystems with the surrounding
large-scale structure (LSS), which indicates the preferential direction of
accretion.

\subsection{The shapes of the galactic subsystems}
\begin{figure*}
  \vspace{-0.4cm}
  \plotone{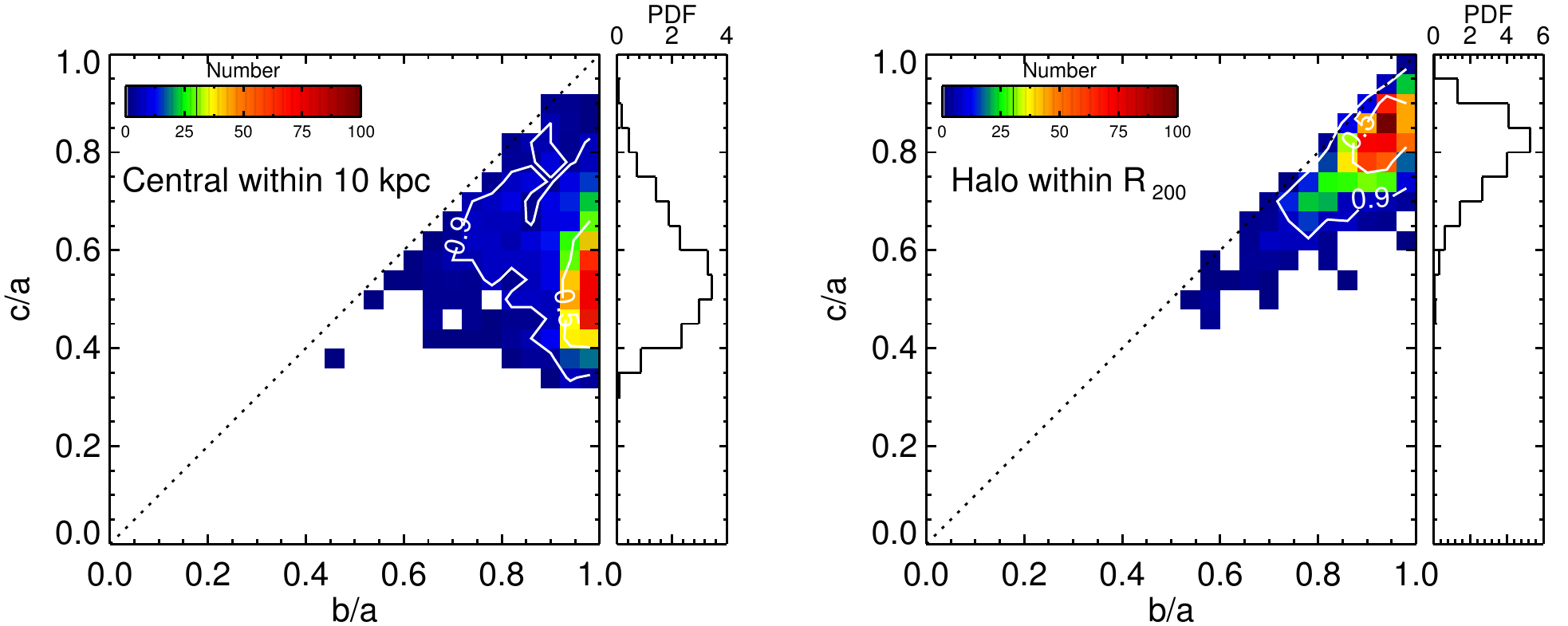}\\[-0.1cm]
  \plotone{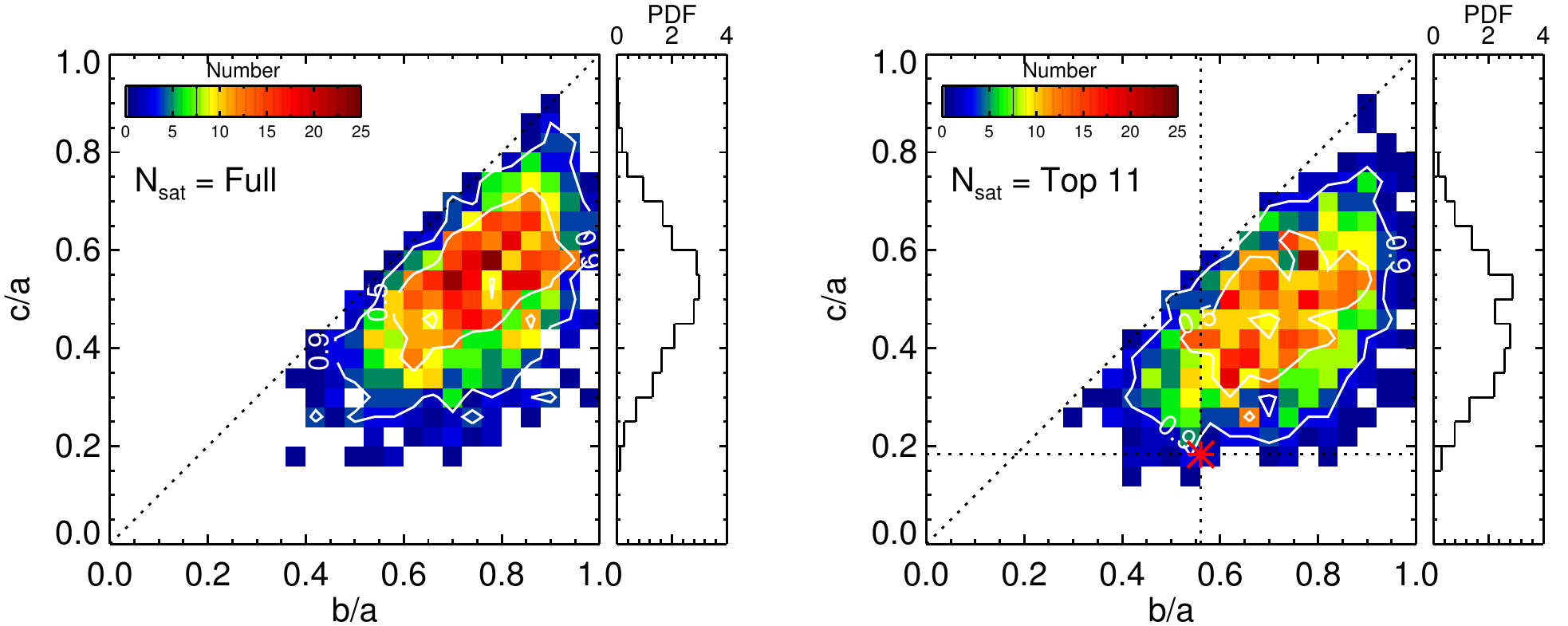}\\[-0.1cm]
  \caption{The distribution of the shape parameters, $b/a$ and $c/a$,
    for central galaxies (top-left), DM haloes (top-right), all the
    luminous satellites (bottom-left) and the 11 most massive satellites,
    ranked by stellar mass (bottom-right). The shapes are computed using
    all the star particles within $10\kpc$ for the central galaxy, all the
    DM particles within $R_{200}$ for the halo, and the luminous
    satellites within a distance of $300\kpc$ from the central galaxy. The
    colours indicate the number of systems in each bin with the
    corresponding numbers given in the top left-hand colour bar. The two solid
    contours indicate the regions enclosing 50 and 90 percent of the
    sample. The right-hand side of each plot shows the PDF of $c/a$.
    The red symbol in the bottom right-hand panel
    shows the axis ratios for the MW's 11 classical satellites.
  }
  \label{fig:dd-caba}
  \vspace{-0.4cm}
\end{figure*}

The shapes of the halo, central galaxy and, to a lesser extent, the
satellite distribution have been studied extensively in both
collisionless and hydrodynamic simulations
\citep[e.g.][]{Bett_10,Wang_13,Tenneti_14,Velliscig_15a}. We therefore
present only a brief overview of the degree of flattening of these
subsystems. \reffig{fig:dd-caba} shows the axes or shape ratios, $b/a$
and $c/a$, for the central galaxy, DM halo and satellite system. The
panels show a two-dimensional histogram where each bin is coloured
according to the number of systems in that bin as indicated by each
colour bar. To the right of each plot, we also show the probability
distribution function (PDF) of $c/a$.

Most central galaxies have a strongly oblate shape ($a \approx b > c
\approx 0.5a$), with more than half of the population having $b/a \geqslant
0.9$ and $0.4 \leqslant c/a \leqslant 0.6$ (see top left-hand panel in
\reffig{fig:dd-caba}). The remaining galaxies are also preferentially
oblate, though to a lesser extent. Note that due to the use of a Jeans
mass limiting pressure floor in the \eagle{} prescription for star
formation, it is difficult for gas to cool into thin discs before
forming stars \citep[for details see][]{Schaye_15}, which results in
an artificial thickening of the stellar component and which may
explain why there are no galaxies with $c/a \leqslant 0.3$. The use of a
mass- rather than light-weighted inertia tensor also leads to thicker
disc.

The DM halo is the closest to spherical of the galactic subsystems
shown in \reffig{fig:dd-caba}, with most haloes having a slightly
prolate ($a > b \approx c$) or nearly spherical ($a \approx b \approx
c$) shape, in agreement with previous studies
\citep[e.g.][]{Frenk1988,Tenneti_14}.

The satellite systems, both for the full and the 11 most massive
objects, have the largest spread in shape parameters, centred on $b/a
\approx 0.7$ and $c/a \approx 0.5$. The total population of subhaloes
is expected approximately to trace the DM halo shape, so the large
spread and the low sphericity ($c/a$) values of the satellite
population reflect systematic effects due to the low number of such
objects \citep{Hoffmann2014} and the biased spatial distribution of
the brightest satellites. This can be appreciated in the lower two
panels of \reffig{fig:dd-caba}, with the system of the 11 most massive
satellites having systematically lower $b/a$ and $c/a$ values than the
full sample of luminous satellites, as noted by \cite{Wang_13}. The
red symbol in the bottom right-hand panel of \reffig{fig:dd-caba} marks the
shape of the 11 classical satellites of the MW, $b/a =0.56 \pm 0.02$
and $c/a = 0.183 \pm 0.008$ (obtained using the positional data from
\citealt{Cautun_15}). While the $b/a$ value for the MW satellites is
typical of the simulated systems, the $c/a$ value is low, with only
$\approx 1\%$ of \eagle{} systems having an equal or lower
sphericity. This is in agreement with previous studies
\citep[e.g.][]{Libeskind_05, Wang_13} that have investigated the high
flattening of the classical MW satellite plane.

\subsection{The alignment of galactic subsystems}
\label{sect:align}
\begin{figure}
  \plotone{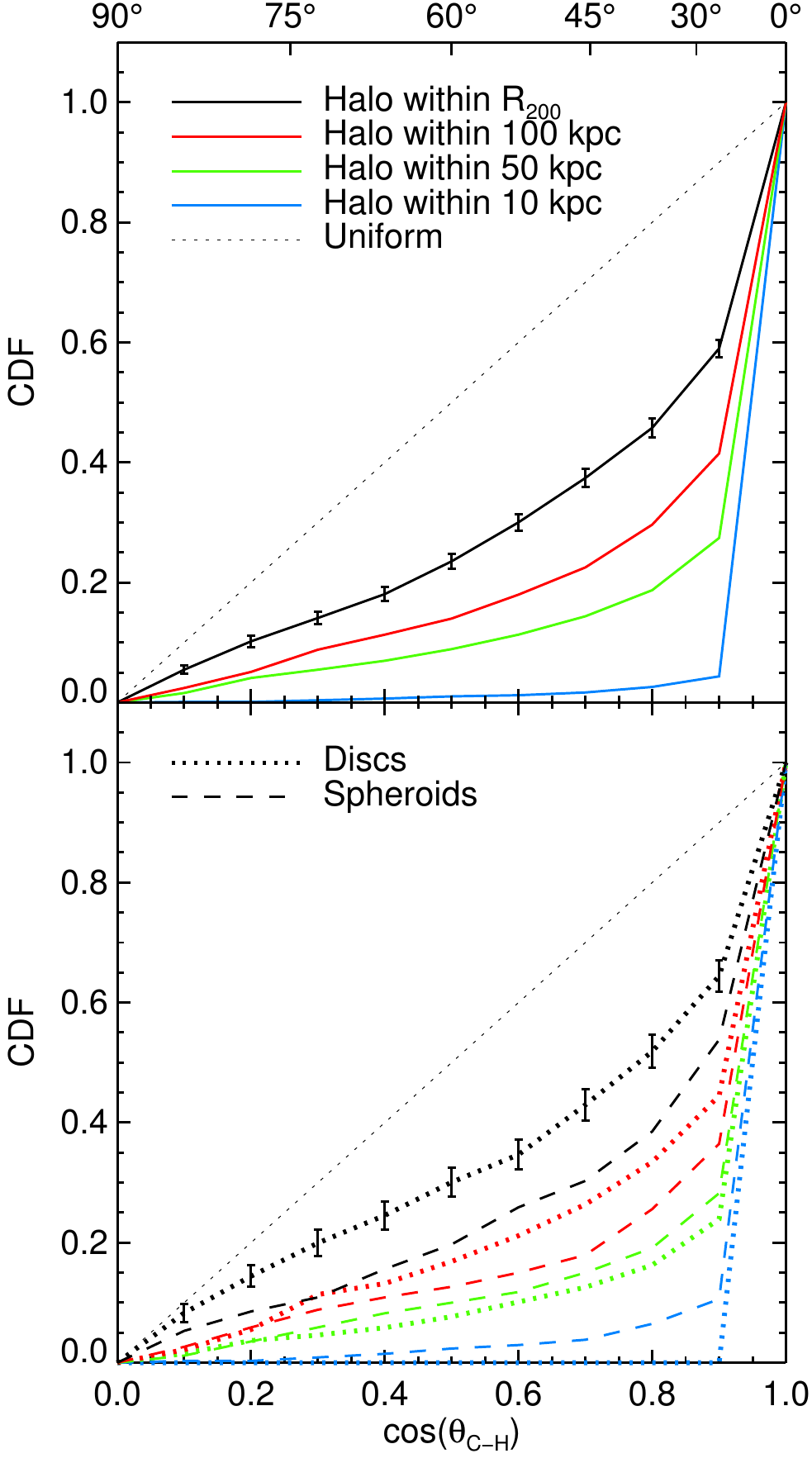}
  \caption{Top panel: the cumulative distribution function (CDF) of
    the misalignment angle, $\cos\theta_{\rm C \td H}$, between the minor axes
    of the central galaxy and the host DM halo. The various lines show the
    dependence of the alignment strength on the region used to determine
    the halo shape, which we measure within spherical regions of radii
    $10, 50, 100\kpc{}$ and $R_{200}$. Bottom panel: same as the top
    panel, but with the central galaxies divided into discs (dotted line)
    and spheroids (dashed line). The error bars show the $1 \sigma$
    bootstrap uncertainty. The thin dotted line in both panels corresponds
    to the CDF of a uniform distribution.
  }
  \label{fig:dis-CH}
  \vspace{-0.4cm}
\end{figure}

We start by studying the alignment between the central galaxy and its
host halo, which we show in the top panel of
Fig.~\ref{fig:dis-CH}. Since the shape and the main axes of the halo
vary as a function of distance from the centre \citep[see
e.g.][]{Bett_10, Velliscig_15a}, we measure the alignment for several
radial extents of $10$, $50$, $100\kpc{}$ and $R_{200}$ by plotting
the cumulative distribution function (CDF) of $\cos\theta_{\rm C \td H}$.
The alignment is the strongest between the innermost halo and the
central galaxy, most likely due to the dominance of baryons in this
inner region, and decreases rapidly as we consider the more extended
halo. The entire halo enclosed within $R_{200}$ still shows a
substantial alignment with the central galaxy, with half of the sample
having a misalignment angle, $\theta_{\rm C \td H} \leqslant 33^\circ$, as shown
in Table~\ref{tab:alignment_angles}.

\begin{table}
  \caption{The misalignment angle, $\theta$ (columns 5-7), and its
    cosine, $\cos\theta$ (columns 2-4), corresponding to $25$, $50$
    and $75$ per cent of the population. Bootstrap resampling gives an
    uncertainty of $\pm0.015$ in the value of $\cos\theta$. The
    corresponding uncertainty for $\theta$ depends on the value of the
    angle and ranges from $\pm3^\circ$ for small angles to $\pm1^\circ$
    for large angles.
  }
  \begin{tabular}{l  c c c  r c c }
    \hline
    Alignment type  & \multicolumn{3}{c}{$\cos\theta$} & \multicolumn{3}{c}{$\theta$ ($^\circ$)} \\
     & ${25\%}$ & ${50\%}$ & ${75\%}$ & \multicolumn{1}{p{.04\textwidth}}{\raggedleft ${25\%}$} & ${50\%}$ & ${75\%}$ \\
    \hline
    Central--halo        & $0.52$ & $0.84$ & $0.96$ & $58$ & $33$ & $17$ \\[.0cm]
    Satellites--halo     & $0.58$ & $0.83$ & $0.95$ & $55$ & $34$ & $19$ \\[.0cm]
    Halo--LSS            & $0.48$ & $0.78$ & $0.93$ & $61$ & $39$ & $22$ \\[.0cm]
    Satellites--LSS      & $0.38$ & $0.66$ & $0.85$ & $68$ & $49$ & $32$ \\[.0cm]
    Central--LSS         & $0.33$ & $0.63$ & $0.85$ & $71$ & $51$ & $32$ \\[.0cm]
    Central--satellites  & $0.30$ & $0.61$ & $0.84$ & $73$ & $52$ & $33$ \\[.0cm]
    Uniform              & $0.25$ & $0.50$ & $0.75$ & $76$ & $60$ & $41$ \\
    \hline
  \end{tabular}
  \label{tab:alignment_angles}
  \vspace{-0.2cm}
\end{table}

Motivated by previous studies which have reported a stronger alignment
for spheroidal galaxies \citep[e.g.][]{Tenneti_15}, we show in the
bottom panel of Fig.~\ref{fig:dis-CH} the misalignment angle,
$\theta_{\rm C \td H}$, separately for disc and spheroid central
galaxies. Because of the limited size of the samples ($\approx 350$
objects each), we assess the significance of any trend with galaxy
morphology using the Kolmogorov-Smirnov (KS) test. The inner
$10\kpc{}$ halo is more aligned for disc galaxies than for the
spheroid population with a KS-test significance of $8.3\sigma$. This
trend reverses as the radial distance used to compute the halo shape
increases such that the entire, $R_{200}$, halo is more aligned with
spheroids than with discs, at a KS significance of $3.7\sigma$. These
results are consistent with observational data \citep[e.g.][for more
details see the discussion section]{Yang_06} and with other
hydrodynamical simulations \cite[e.g.][]{Tenneti_15}, but are contrary
to the results of \citet{Velliscig_15a}, which found that disc
galaxies are better aligned with their haloes than spheroidal
ones. The discrepancy is due to the method used to classify the
galaxies into discs and spheroids.
\citet{Velliscig_15a} used the ratio of the \textsc{subfind} velocity
dispersion to the maximum circular velocity, while we used the fraction
of the kinetic energy in ordered rotation that, with hindsight,
is a better kinematical indicator of galaxy morphology.

\begin{figure}
  \plotone{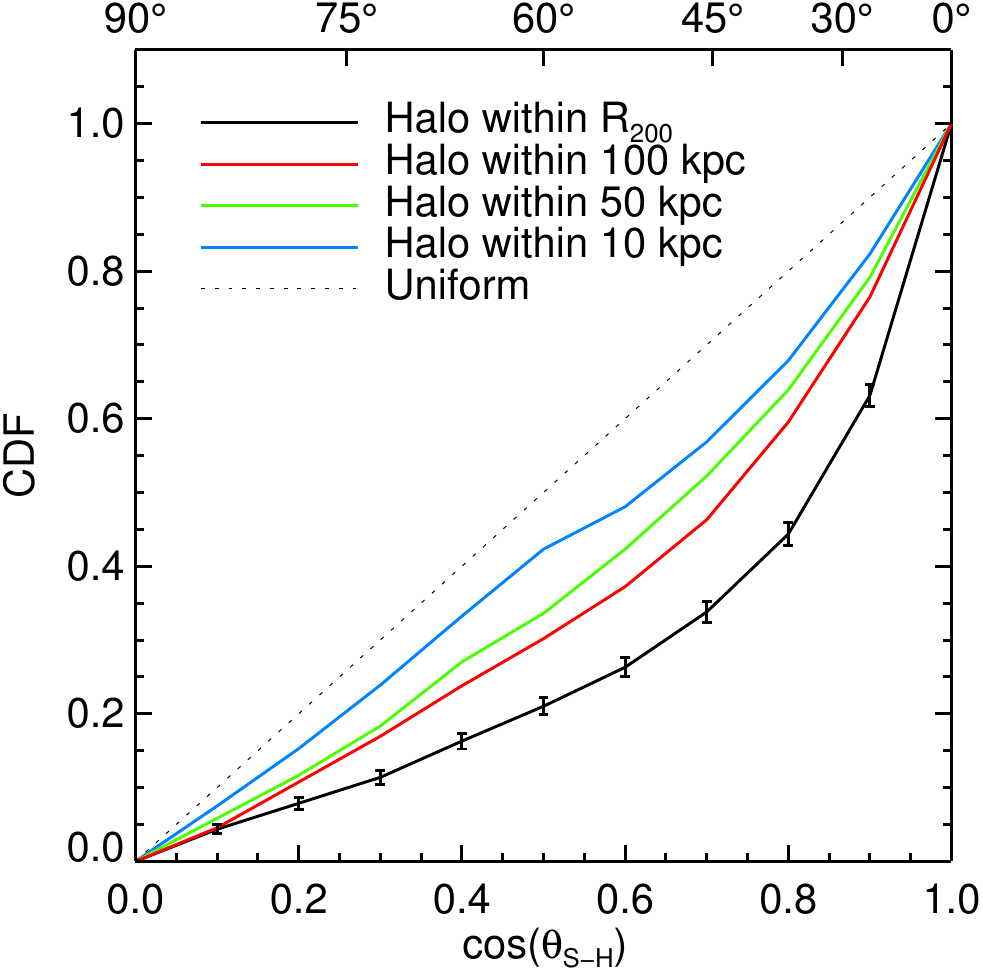}
  \caption{The CDF of the misalignment angle, $\cos\theta_{\rm S \td H}$,
    between the minor axes of the satellite system and the host DM halo,
    with the halo shape measured within various radial distances.
  }
  \label{fig:dis-SH}
  \vspace{-0.4cm}
\end{figure}

In Fig.~\ref{fig:dis-SH}, we compare the alignment between the
satellite system and its host halo, again with the halo shape
measured as function of distance from the centre. In contrast to the
central--halo alignment, the satellites are more aligned with the
entire halo and to a much lesser extent with the inner regions of the
halo. This is to be expected, since the satellite system is more
extended than the central galaxy and is thus more likely to trace the
outer halo.

\begin{figure} 
  \plotone{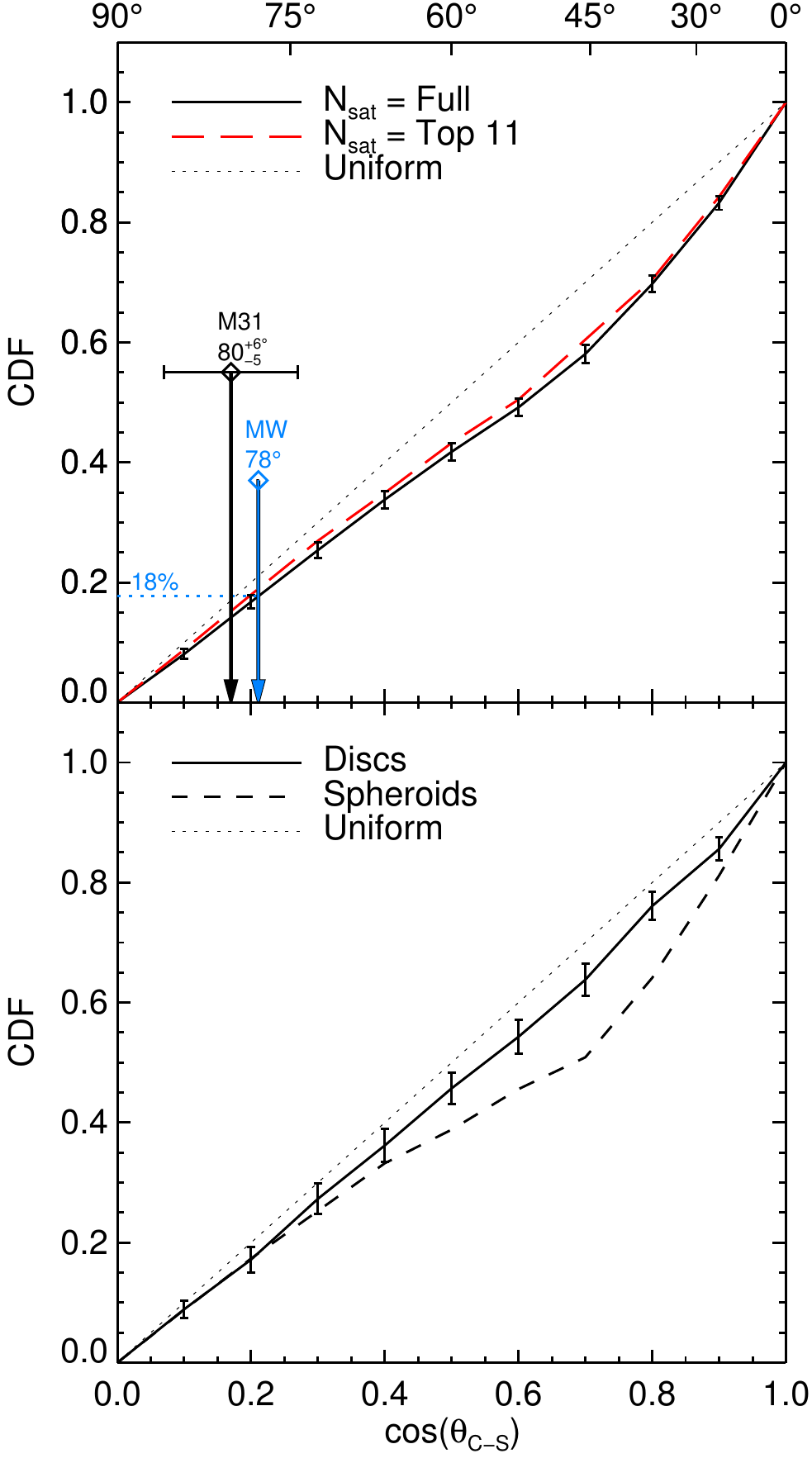}
  \caption{The CDF of the
    misalignment angle, $\cos\theta_{\rm C \td S}$, between the minor axes of
    the central galaxy and the satellite system. Top panel: the
    solid line indicates the alignment of the full set of satellites while
    the dashed line shows the alignment of the most massive 11
    satellites. The right vertical arrow shows the misalignment angle,
    $\theta_{\rm C \td S;\; MW}= 78^\circ$, for the MW system while the left
    vertical arrow with error bars shows the misalignment angle,
    $\theta_{\rm C \td S;\; M31}={80^\circ}_{-5}^{+6}{\;}$, and its $1\sigma$
    range for the M31 system. Bottom panel: the alignment of the full set
    of satellites with disc (solid) and spheroid (dashed) central
    galaxies.
  }
  \label{fig:dis-CS}
  \vspace{-0.4cm}
\end{figure}

Fig.~\ref{fig:dis-CS} shows that there is an alignment, albeit weak,
between the central galaxies and their satellite systems. The
alignment strength is very similar if we consider the full set of
luminous satellites in \eagle{} (solid line) or only the 11 most
massive satellites (dashed line), which would correspond to the
classical satellites of the MW. The bottom panel of
Fig.~\ref{fig:dis-CS} shows that spheroid centrals are more aligned
with their satellite systems than disc centrals, though in both cases
the alignment is weak. The dependence of alignment on central
morphology is robust, having a KS test significance of $3.0\sigma$.

The alignment between central galaxies and their satellites, as
measured in \eagle{}, is important to better understand the two major
satellite systems in the LG. For this, we computed the misalignment
angle, $\theta_{\rm C \td S}$, of the MW and M31 systems, whose values are
shown by two vertical arrows in the top panel of
Fig.~\ref{fig:dis-CS}. In the case of the MW, we considered only the
11 classical satellites, since observations of fainter satellites are
more strongly affected by incomplete survey area and
incompleteness. Using the coordinates and uncertainties from
\citet{McConnachie2012}, we computed a misalignment angle,
$\theta_{\rm C \td S;\; MW}= 78^\circ$, $\cos\theta_{\rm C \td S;\; MW}=0.21
\pm 0.01$, between the disc of the MW and its 11 classical
satellites. In the case of M31, using the \citet{McConnachie2012}
catalogue, we selected as satellite galaxies brighter than $-8.8$ in
absolute \textit{V}-band magnitude (equal to the faintest classical MW
satellite) that is within a 3D-distance of $300\kpc$ from M31. This
resulted in $18$ satellites whose spatial distribution has axis
ratios, $b/a = 0.72_{- 0.06}^{+0.07}$ and $c/a =
0.61_{-0.04}^{+0.03}$, and has a misalignment angle, $\theta_{\rm
C \td S;\; M31}= {80^\circ}_{-5}^{+6}{\;}$, i.e. $\cos\theta_{\rm C \td S;\;
M31}=0.17_{-0.10}^{+0.09}$, with the disc of M31. We quote $1\sigma$
uncertainties due to errors in the distance of the M31 satellites.
Thus, both the MW and M31 have systems of bright satellites that are
nearly perpendicular to their disc. Such configurations are quite
common, with $\approx 20\%$ of the \eagle{} systems having
misalignment angles at least as extreme as the MW and M31.  A notable
feature of the classical MW satellites is that they are distributed
along a thin plane, so we checked if the central--satellite system
alignment is correlated to the shape of the satellite distribution and
find no such dependence.  Fig.~\ref{fig:P1} shows a selection of five
such systems, i.e. with $\theta_{\rm C \td S} \geqslant 78^\circ$. Each
panel shows two perpendicular views of the distributions of stars and
satellites in those haloes. Some of these systems, like those shown in
the middle three panels, have thin satellite planes,
i.e. $c/a\sim0.2$, that are also nearly perpendicular on their central
galaxy, as is the case for our own Galaxy.

\subsection{Conditional alignments: the key to a better understanding}
\label{sect:conditional_align}

\begin{figure}
  \plotone{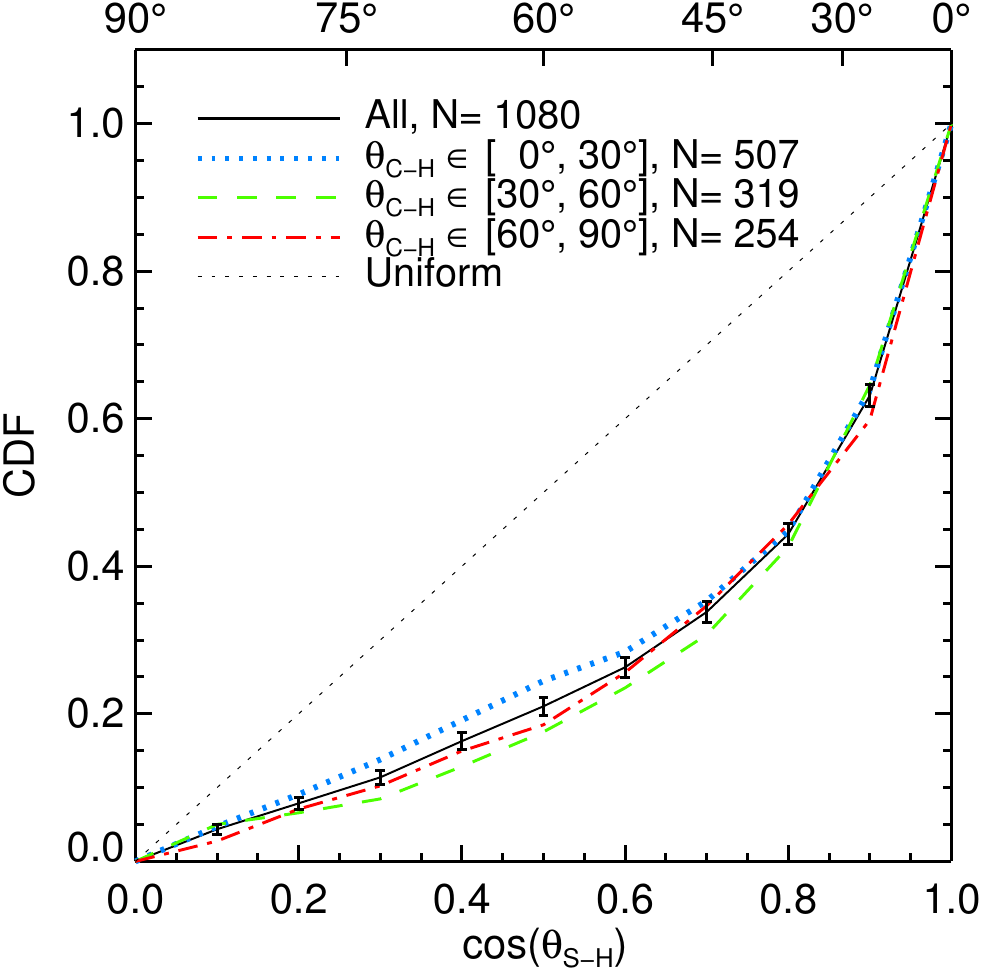}
  \caption{The \textit{conditional} alignment, $\cos\theta_{\rm S \td H}$,
    between satellite systems and their DM haloes given the misalignment angle,
    $\theta_{\rm C \td H}$, between the centrals and their DM haloes. We show
    the alignment of the full sample (solid line) and that of various
    subsamples selected according to the value of $\theta_{\rm C \td H}$. The
    subsamples correspond to the central and halo being: aligned,
    $0^{\circ} \leqslant \theta_{\rm C \td H} \leqslant 30^{\circ}$ (dotted line);
    intermediate, $30^{\circ}< \theta_{\rm C \td H} \leqslant 60^{\circ}$ (dashed
    line); and perpendicular, $60^{\circ}< \theta_{\rm C \td H} \leqslant 90^{\circ}$
    (dashed-dotted line). See the plot legend for the number of systems in
    each subsample.
  }
  \label{fig:eagle-CH-SH}
\end{figure}

\begin{figure}
  \plotone{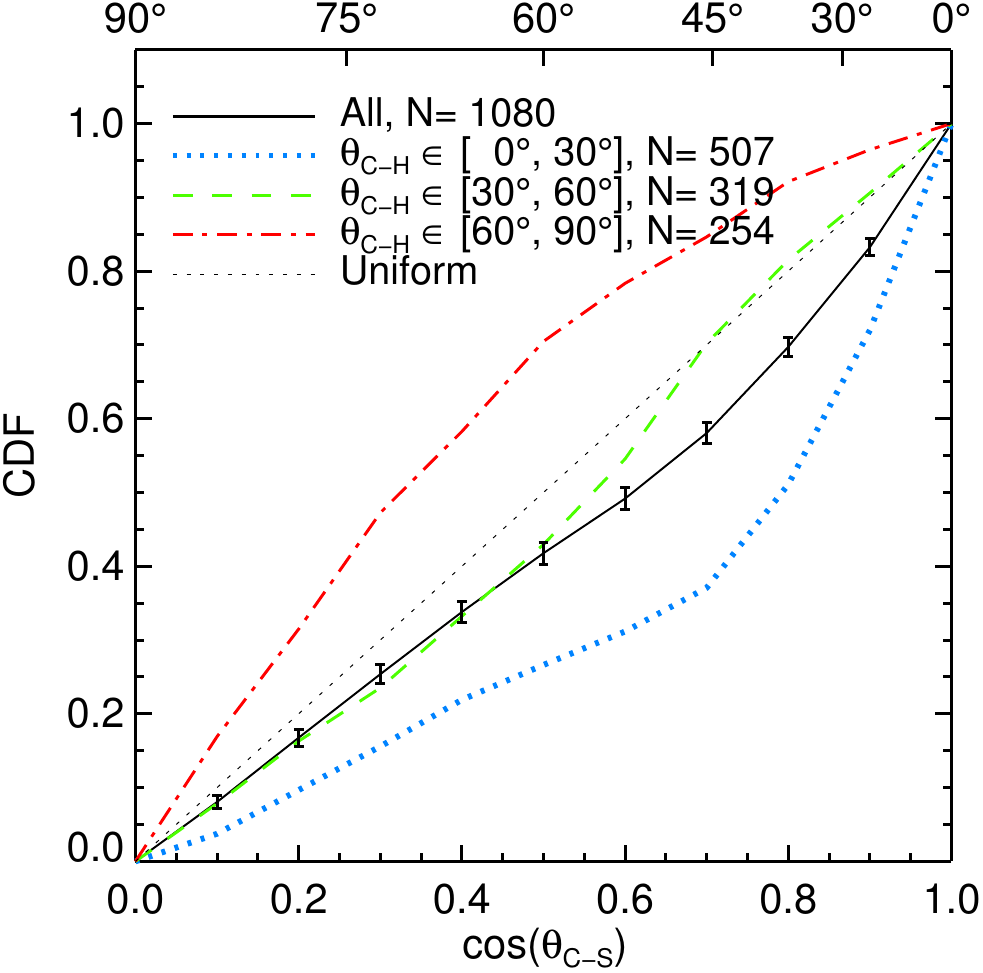}
  \caption{Same as Fig.~\ref{fig:eagle-CH-SH}, but for the
    \textit{conditional} alignment, $\cos\theta_{\rm C \td S}$, between
    centrals and their satellite systems given the misalignment angle,
    $\theta_{\rm C \td H}$, between the centrals and their DM haloes.
  }
  \label{fig:eagle-SH-CS}
\end{figure}

\begin{figure}
  \plotone{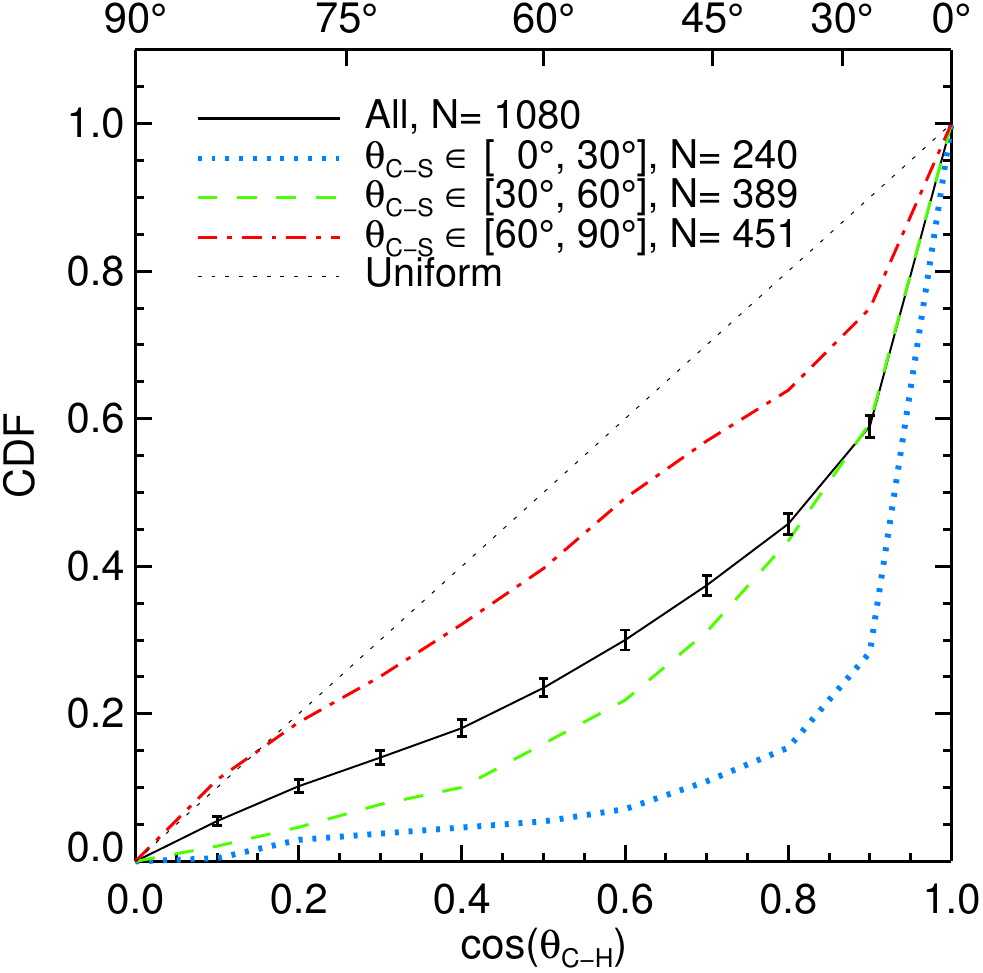}
  \caption{Same as in Fig.~\ref{fig:eagle-CH-SH}, but for the
    \textit{conditional} alignment, $\cos\theta_{\rm C \td H}$, between
    centrals and their DM haloes given the misalignment angle, $\theta_{\rm
      C \td S}$, between the centrals and their satellite systems.
  }
  \label{fig:eagle-CS-CH}
\end{figure}

We found that both the central galaxies and the satellite systems have
a strong alignment with a third component, the DM halo. This naturally
gives rise to an indirect, or secondary, alignment between the central
galaxies and their satellites since both are aligned with their DM haloes.
In the following, we wish to investigate if this effect can explain the
weak alignment between centrals and their satellite systems. We do so
by studying \textit{conditional} alignments, that is, alignments of a
subsample of objects that satisfy a certain condition.

If the central--satellite alignment is not just a byproduct of the
alignment of both components with the halo we would expect a stronger
alignment for systems in which the halo and the central are aligned.
This arises because the satellites will feel the combined coherent
effect of being aligned with both the halo and the central
galaxy. This effect is studied in Fig.~\ref{fig:eagle-CH-SH} where we
show the satellite--halo alignment \textit{conditional} on the central--halo
misalignment angle, $\theta_{\rm C \td H}$. We split our sample into three
subsamples according to the value of $\theta_{\rm C \td H}$ as follows:
aligned, $\theta_{\rm C \td H} \in [0^{\circ}, 30^{\circ}]$; intermediate,
$\theta_{\rm C \td H} \in [30^{\circ}, 60^{\circ}]$; and perpendicular,
$\theta_{\rm C \td H} \in [60^{\circ}, 90^{\circ}]$. As
Fig.~\ref{fig:eagle-CH-SH} shows, all three subsamples have the same
degree of alignment between the satellites and the halo as the overall
sample, suggesting that the central galaxy does not directly influence
the orientation of the satellite system.

Fig.~\ref{fig:eagle-SH-CS} shows a complementary test where, using the
same subsamples as in Fig.~\ref{fig:eagle-CH-SH}, we show the
\textit{conditional} alignment between centrals and their
satellites. The misalignment degree varies vastly between subsamples:
the centrals that are more aligned with their haloes are also the ones
that are more aligned with their satellite systems.

To summarize, the central--satellite alignment is a consequence of
the tendency of both components to align with the halo. This result
has important applications since it can be used to predict with some
confidence the orientation of the DM halo from the orientation of its
galaxies only, as illustrated in Fig.~\ref{fig:eagle-CS-CH}. The
figure shows that if the central and the satellite system are aligned
($\theta_{\rm C \td S} \leqslant 30^\circ$), then the DM halo system also
tends to point to the same direction. In contrast, if the satellite
system is perpendicular to the disc of the central ($\theta_{\rm C \td S}
\geqslant 60^\circ$), as in the case of the MW and M31, then the DM
halo is only poorly aligned with the central. The dependence on
$\theta_{\rm C \td S}$ is strong, with the median central--halo
misalignment angle, which is $33^\circ$ for the entire sample, varying
from $18^\circ$ for $\theta_{\rm C \td S} \leqslant 30^\circ$ to
$52^\circ$ for $\theta_{\rm C \td S} \geqslant 60^\circ$.

\subsection{The alignment with the large-scale structure}
\label{sect:LSS_align}

\begin{figure}
  \plotone{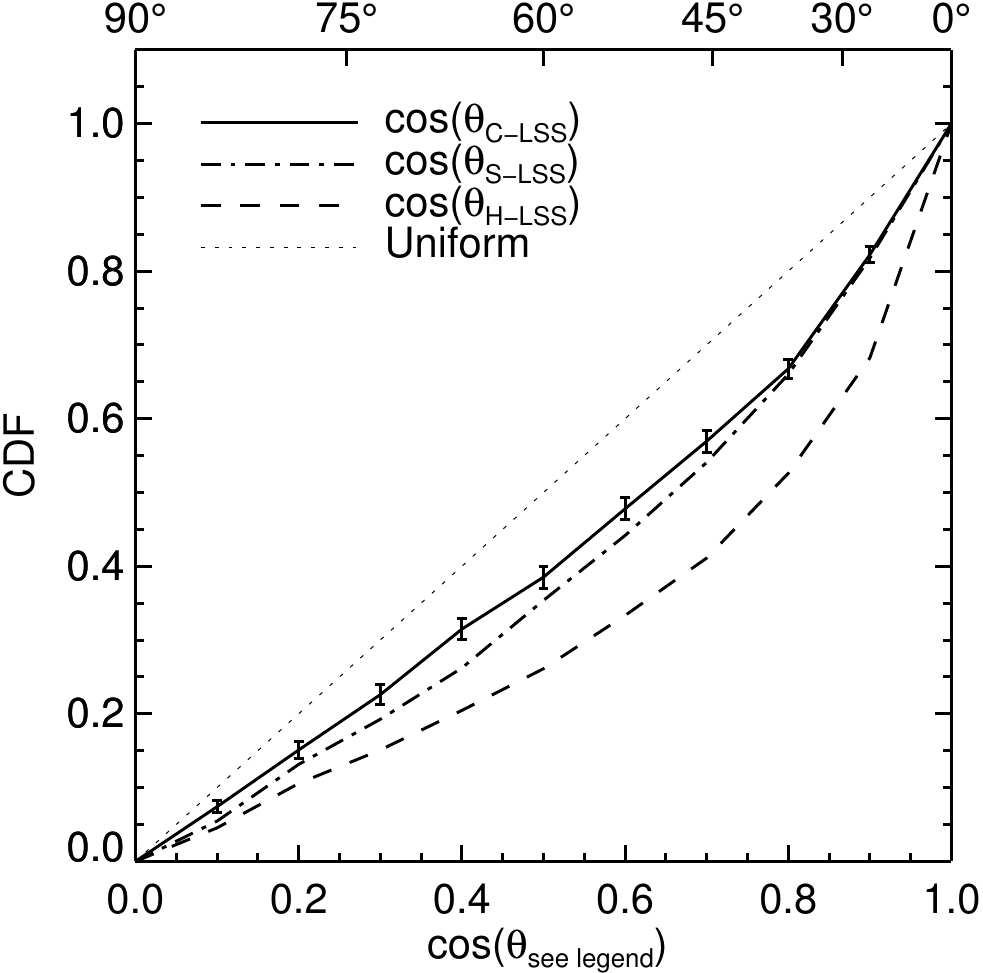}
  \caption{The CDF of the
    misalignment angles of the various galactic subsystems with the LSS in
    which they are embedded (on scales of $2~\td~3R_{200}$). The solid line
    shows the central galaxy--LSS alignment, $\cos\theta_{\rm C \td LSS}$;
    the dashed-dotted line the satellite system--LSS alignment,
    $\cos\theta_{\rm S \td LSS}$; and the dashed line the halo--LSS
    alignment, $\cos\theta_{\rm H \td LSS}$.
  }
  \label{fig:dis-lss}
\end{figure}

\begin{figure}
  \plotone{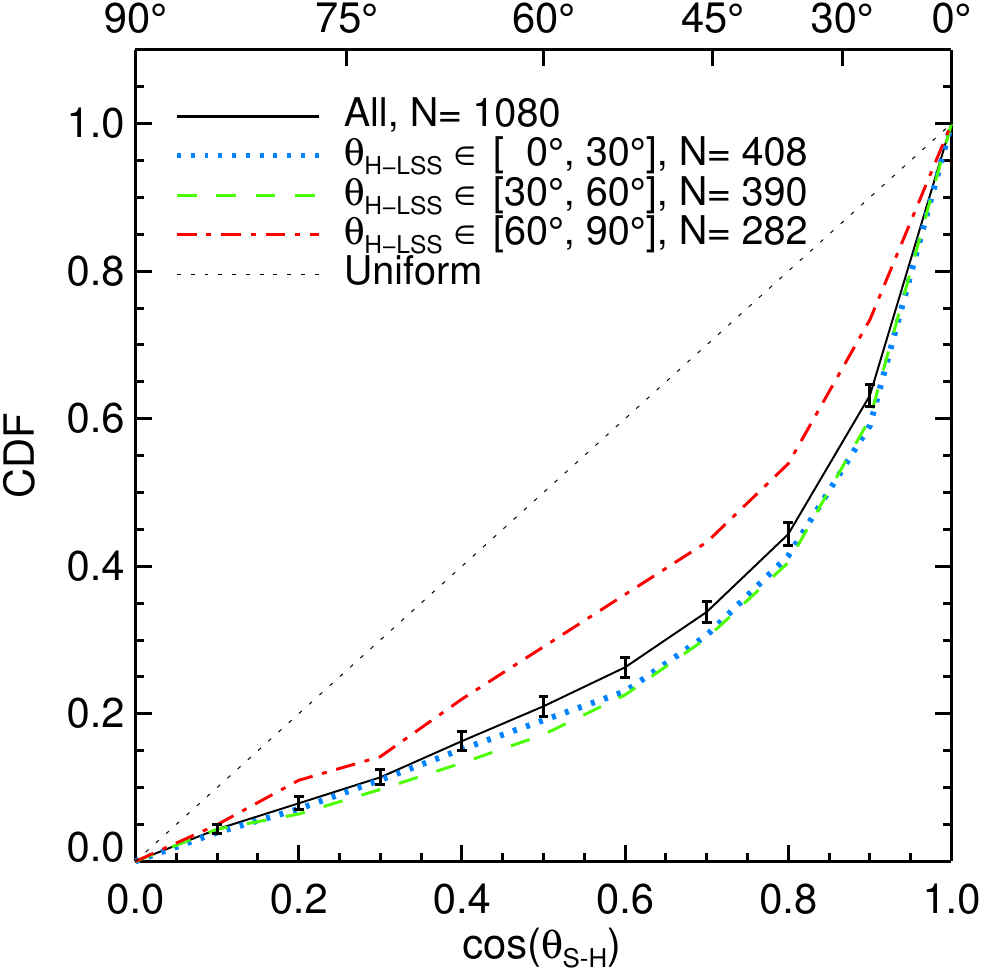}
  \caption{Same as Fig.~\ref{fig:eagle-CH-SH}, but for the
    \textit{conditional} alignment, $\cos\theta_{\rm S \td H}$, between
    satellite systems and their DM halo given the misalignment angle,
    $\theta_{\rm H \td LSS}$, between the DM haloes and the LSS within which they are embedded
    (on scales of $2~\td~3R_{200}$).
  }
  \label{fig:dis-lss-split}
\end{figure}

Within the standard model, DM, gas and satellites are accreted
predominantly along filaments, which determine a common preferred axis
\citep[e.g.][]{Libeskind_05,Libeskind2011,Libeskind2014,Deason2011,Lovell2011}. Thus,
we would expect the halo, central and satellite system to be aligned
with the LSS in which they are embedded
\citep[e.g.][]{Tempel_15,Velliscig_15b,Welker2015}.

We measure the orientation of the LSS by computing the moment of
inertia of the matter within the spherical shell located between
$2R_{200}$ and $3R_{200}$ from the centre of each halo. We then
compute the misalignment angle between the minor axes of the galactic
subsystems and that of the LSS. The resulting alignment is shown in
Fig.~\ref{fig:dis-lss}. We find that all three galactic components
show some degree of alignment with their surrounding distribution of
matter: the halo--LSS alignment is the largest, followed by the
satellite--LSS and central--LSS alignments (see
Table~\ref{tab:alignment_angles} for a comparison to the alignment
between galactic components). We note that the alignment with the LSS
decreases rapidly if we were to measure the LSS directions using
spherical shells of larger radii.

In Section~\ref{sect:conditional_align}, we found that the central--satellites
alignment is a consequence of both components being aligned
with a third, the DM halo. Since the LSS shows a considerable
alignment with the halo, we studied if the satellite--LSS and central--LSS
alignments are a consequence of the same effect. The former is
investigated in Fig.~\ref{fig:dis-lss-split}, where we show the
satellite--halo alignment for subsamples selected according to the
halo--LSS misalignment angle, $\theta_{\rm H \td LSS}$. We find that the
satellite--halo alignment is weaker for higher values of $\theta_{\rm
H \td LSS}$, i.e. when the halo is close to perpendicular to the
LSS. Thus, the satellite system is more strongly aligned with the LSS
than would be expected from the fact that both are aligned with the
halo. In contrast, the central--LSS alignment is a consequence of the
tendency of both components to be aligned with the halo. Applying the
same test as in Fig.~\ref{fig:dis-lss-split} to the central--halo
alignment, we found no significant trend with $\theta_{\rm H \td LSS}$.

\section{Discussion}
\label{sect:disc}
We have studied the alignment between the central galaxy, satellite
system and DM halo as well as that of the LSS within which they are
embedded. The sample consists of 1080 MW-mass systems (of typical mass
$\sim 10^{12}\Msun$) that have at least 11 luminous satellites within
$300\kpc$, similar to the MW and M31 systems. This sample was selected
from the largest of the \eagle{} hydrodynamical simulation, which is
an ideal tool for our study. First, the \eagle{} simulation has been
calibrated to reproduce the observed galaxy stellar mass function and
the observed size-mass relation \citep{Crain_15,Schaye_15}. Secondly,
the resolution of \eagle{} is sufficient to identify luminous
satellites that are comparable to the classical dwarf satellites of
the MW while providing a large enough sample of MW-mass haloes. In the
following, we discuss the major results of this work.

\subsection{Alignments with the DM halo}
We find that central galaxies tend to be well aligned with their DM
host haloes, with a median misalignment angle of $33^\circ$, which is
in good agreement with previous studies
\citep[e.g.][]{Bett_10,Tenneti_14,Velliscig_15a}. The centrals show an
even larger degree of alignment with the inner halo
\citep{Bett_10,Deason2011,Velliscig_15a}, with most centrals being
nearly parallel to the halo orientation within $50\kpc$ or
less. \citet{Bailin_05} found the same result and argued that in the
inner $\sim 20\kpc$ region the baryonic and DM components exert a
similar torque on each other and thus are equally responsible for
their very strong alignment.

While the centrals tend to be very well aligned with the inner
$10\kpc$ halo, \reffig{fig:dis-CH} also shows that this inner $10\kpc$
halo is only partially aligned with the outer halo, with a median
misalignment angle of $33^\circ$. This misalignment is stronger than
that measured in DM-only simulations, with \citet{Bailin2005}
reporting a median misalignment angle of $\approx 25^\circ$. The increased
misalignment is likely due to the presence of baryons that affect the
orientation of the inner halo while hardly affecting the outer halo
\citep{Bailin_05}.

\begin{figure}
  \plotone{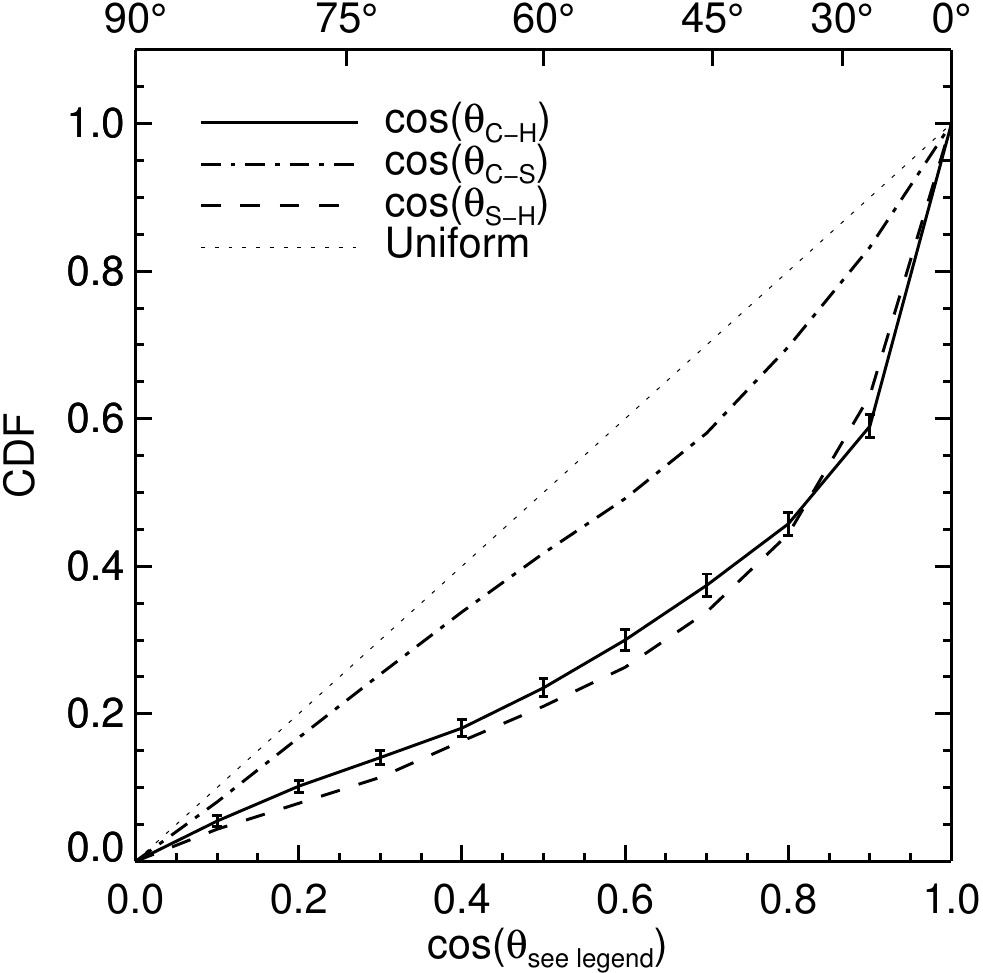}
  \caption{A summary of the alignment between the central galaxies
    (C), their satellite systems (S), and their entire DM haloes (H),
    i.e. within $R_{200}$. The plot shows the CDF of the misalignment angle,
    $\theta$, between: centrals and their haloes (solid line), satellite
    systems and their haloes (dashed line), and centrals and their
    satellites (dash-dotted line).
  }
  \vspace{-0.2cm}
  \label{fig:dis-CSH}
\end{figure}

The satellite system is misaligned with the entire halo to the same
extent as the central galaxy is, as can be seen from
Fig.~\ref{fig:dis-CSH}. This is somewhat surprising, since many
satellites are found in the outer regions of the halo, and may thus be
expected to trace the entire halo quite well. Note that we do find
that the satellites are more misaligned with the inner halo than with
the entire halo and thus they do preferentially trace the outer
halo. This misalignment between satellites and their host halo can be
attributed to two causes. First, the misalignment is partially due
to the relatively small number of satellites per halo (on average
$15$), which means that Poisson noise plays an important role
\citep{Hoffmann2014}. Secondly, the accretion of the most massive
satellites is more anisotropic than that of the components that
provide the bulk of the mass of the halo, which are lower mass
subhaloes and smooth accretion \citep{Libeskind2014}. This can lead to
intrinsic differences between the luminous satellites and the DM halo.

Individually, neither the central galaxy nor the satellite population
are very good predictors of the orientation of the DM halo. But, by
combining the two components, we can find a subsample that shows a
much smaller misalignment angle with the halo. This subsample consists
of systems in which the central is nearly parallel to the satellite
system, for which the median central--halo misalignment angle is just
$18^\circ$. Using a sample with these characteristics would greatly
improve the ability to measure the mean flattening of the DM halo
using stacked lensing maps \citep[see][and references
therein]{Bett2012}, which currently is limited due to the broad
distribution of central--halo misalignment angles. Since 3D satellite
positions are currently restricted to the nearby Universe, more work
is needed to understand if a similar relation would hold when using
projected satellite distributions. Similarly to the central--satellite
alignment, the central--LSS alignment is also a consequence
of the tendency of both components to align with the halo. So,
potentially, to obtain a stronger central--halo alignment, one could
also select systems in which the central is nearly parallel to the LSS.

\subsection{The central--satellite system alignment}
We also studied the alignment between the central galaxy and the
orientation of the entire system of satellites. This is different from
most other studies, which measured the alignment between the position
of individual satellites and the preferential axes of their
central. We found a weak central--satellite system alignment that
suggests that satellites are somewhat more likely to be found along
the plane of the central galaxy \citep[see also][]{Velliscig_15b}.
This is in agreement with \citet[][see also
\citealt{Sales_04,Brainerd_05,Wang_08,Wang_10}]{Yang_06} who measured the same
tendency in observational data. More interestingly, we found that the
central--satellite system alignment is a byproduct of the tendency of both
components to align with the halo \citep[see also][]{Agustsson2010,Wang_13}.

\subsection{Dependence on galaxy type}
We classified our sample into disc (rotating) and spheroid
(non-rotating) subsamples based on the fraction of the kinematic
energy in ordered rotation of each central galaxy, as described in
Section~\ref{sect:kappa}. We found that spheroidal galaxies tend to be
more aligned with both their haloes and their satellite systems. This
difference in alignment strength between spheroids and discs is
statistically significant at the more than $3\sigma$
level. Surprisingly, the inner halo shows the opposite trend, being
more strongly aligned with discs than with spheroids, with an
$8\sigma$ significance.

Compared to observations, if we refer to discs as blue (late type)
galaxies and to spheroids as red (early type) galaxies, our result is
in agreement with the findings of \citet{Yang_06}: red centrals show a
strong alignment with their satellites, while blue centrals have
roughly isotropically distributed satellites. The cause of the trend
in alignment strength with the morphology of the central is a topic of
debate \citep[e.g. see][]{Yang_06,Kang_07}. At least for our sample,
we checked that this trend is not due to spheroidal centrals being
located in more massive haloes than disc centrals. The trend in
alignment strength could be related to the properties of the central
galaxies themselves, e.g. discs, due to their higher specific angular
momentum, are harder to torque than spheroids.

\subsection{The connection to the large-scale distribution of matter}
\label{sect:discussion_LSS}
We found that the halo, central and satellite system tend to align
with the LSS in which they are embedded (on scales of $2~\td~3R_{200}$),
with the former showing the strongest alignment. This agrees with
observational studies that also found that both central and satellite
galaxies are aligned with the preferential directions of the cosmic
web \citep{Paz2011,Zhang_13,Libeskind2015,Tempel_15}. We also found
that the central--LSS alignment is a consequence of both aligning
with the DM halo. In turn, this results in spheroid galaxies being
slightly better aligned with the LSS than discs since spheroids are
more aligned with their host haloes. This trend with central
morphology is seen in observations too, with \citet{Zhang_13} finding
that red centrals are more strongly aligned with the cosmic web than
blue centrals.

The absolute strength of the alignment with the LSS depends on the
scale used to determine the LSS orientation, with a larger scale
resulting in a smaller alignment. In this paper, we used the mass
distribution between two and three times the virial radius, $R_{200}$,
of each halo, which corresponds to scales of $0.5~\td~1\Mpc{}$ to define
the LSS. These scales are considerably smaller than those available to
observations, which are typically a few$\Mpc$. Thus, while we find the
same qualitative results as previous studies, from a quantitative
perspective we have stronger alignments between the galactic
subsystems and the LSS.

\subsection{Implications for the MW and M31}
The \eagle{} simulation indicates that configurations similar to the
MW and M31, in which the satellite population is nearly perpendicular
to the central disc, are quite common. This result was hinted by the
hydrodynamical simulations of \citet{Libeskind2007} who found one such
perpendicular configuration in their sample of just three galaxies. In
fact, $\approx 20\%$ of systems have a misalignment angle larger than
the MW or M31, which have $\theta_{\rm C \td S;\; MW}= 78^\circ$ and
$\theta_{\rm C \td S;\; M31}= {80^\circ}_{-5}^{+6}{\;}$, respectively. This
large fraction of perpendicular systems is due to the weak alignment
between centrals and their satellite systems, which is close to a
uniform distribution.

We also predict that the minor axes of the inner $\sim 10\kpc$ haloes
should be parallel to the normal of the disc planes of the MW and M31,
since the alignment for disc galaxies is very strong. In contrast, the
outer halo should be only weakly aligned with the central galaxy
since, as may be seen in Fig.~\ref{fig:eagle-CS-CH}, the satellite
systems of the MW and M31 are nearly perpendicular to their central
galaxies. We therefore expect the orientation and shape of the halo in
these galaxies to vary significantly with radius, a feature that
should be taken into account when modelling, for example, the dynamics
of MW streams and halo stars \citep[for details see][]{Vera-Ciro2013}.

Recently, \citet{Libeskind2015} analysed the alignment with the LSS of
several nearby satellite planes: the one in the MW, the two in M31
\citep{Ibata_13,Shaya2013} and the two in the Centaurus A Group
\citep{Tully2015}. They found that four out of the five planes,
i.e. all except the one in the MW, are nearly parallel (largest
misalignment angle is $\approx 14^\circ$) to the minor axis of the
cosmic web. Such a result is surprising, since we found a median
satellite system--LSS misalignment angle of $49^\circ$, which is
likely to be much higher when determining the LSS orientation on a
$2.5\Mpc{}$ scale, as done by Libeskind et al. (see the discussion in
Section~\ref{sect:discussion_LSS}). The strong alignment of these
satellite planes with the LSS could be due to the particular
environment of the LG and its immediate neighbourhood which may not be
representative of the Universe as a whole. Alternatively, it may not
be appropriate to compare our results with those of \citet{Libeskind2015},
since their planes consist of subpopulations of satellites that form
spatially thin configurations \citep[for details see][]{Cautun_15} and
not of the entire satellite populations, as we have considered in this
study. Further work is needed to clarify the puzzling alignments
detected by \cite{Libeskind2015} between satellite planes and the
cosmic web.

\vspace{-0.2cm}
\section{Conclusions}
\label{sect:conclusion}
We have studied the alignments of the central galaxy, DM halo and
satellite system at the present-day in the \eagle{} hydro-cosmological
simulation. \eagle{} self-consistently incorporates the main physical
processes that affect galaxy and halo shapes as well as the orbits of
satellite galaxies, and is therefore ideal for our study. Our sample
consists of MW-mass haloes (of typical mass $\sim 10^{12}\Msun$) that
have at least 11 luminous satellites within a radius of $300\kpc$; we
found 1080 such systems in \eagle{}. The main axes were determined
from the moment of inertia measured within $10\kpc{}$ for centrals,
$R_{200}$ for haloes and $300\kpc{}$ for the satellite system. We
focused on the misalignment angle between the minor axes of the
galactic components since the major and intermediate axes show a
lesser degree of alignment.

\noindent ~~~~Our main conclusions are as follows.
\\[-.55cm]
\begin{enumerate}
    \item The central galaxies and the satellite systems tend to
      be well aligned with their host haloes, with a median misalignment
      angle of $\approx 33^\circ$ in both cases (see
      Table~\ref{tab:alignment_angles}). On the other hand, the centrals and
      their satellites are only weakly aligned with one another (see
      Fig.~\ref{fig:dis-CSH}).
      \\[-.3cm]
    \item The alignment strength depends on the radial extent of the
      DM halo considered. The alignment of central galaxies is largest with
      the inner $10\kpc$ of the halo and decreases with increasing radial
      extent (see Fig.~\ref{fig:dis-CH}). In contrast, the satellite system
      is better aligned with the entire halo, as measured within $R_{200}$,
      and less well aligned with the inner halo (see Fig.~\ref{fig:dis-SH}).
      \\[-.3cm]
    \item Spheroidal centrals are better aligned with both their halo
      and their satellite system than disc centrals (see
      Figs.~\ref{fig:dis-CH}~and~\ref{fig:dis-CS}).
      \\[-.3cm]
    \item The weak alignment between centrals and their satellites is
      a consequence of the tendency of both components to be aligned with
      the DM halo (see
      Figs.~\ref{fig:eagle-CH-SH}~and~\ref{fig:eagle-SH-CS}).
      \\[-.3cm]
    \item The orientation of the halo can be tightly constrained in
      systems where the centrals and satellite systems are close to
      parallel, with such subsamples having a median central--halo
      misalignment angle of only $18^\circ$. In contrast, systems where the
      central and satellite systems are nearly perpendicular, as is the case
      for the MW and M31, show a much weaker central--halo alignment (see
      Fig.~\ref{fig:eagle-CS-CH}).
      \\[-.3cm]
    \item The central, halo and satellites tend to be aligned, to
      various degrees, with the large-scale distribution of matter in which
      they are embedded (see Fig.~\ref{fig:dis-lss}). While the central--LSS
      alignment is a consequence of both components being somewhat
      aligned with the halo, the satellite--LSS alignment is stronger than
      expected from such an effect alone (see Fig.~\ref{fig:dis-lss-split}).
\end{enumerate}

To conclude, our goal was to better understand the seemingly puzzling
situation around the MW and M31 where the configurations of bright
satellites are nearly perpendicular to the disc of their centrals.
Because of the weak alignment between centrals and their satellites, such
perpendicular configurations are in fact quite common, with $\approx
20\%$ of \eagle{} systems having misalignment angles at least as
extreme as the MW and M31. The perpendicular configuration also
implies that the directions of the MW and M31 haloes cannot be
constrained from the orientation of their centrals, since such systems
have only a very weak central--halo alignment.

\vspace{-0.2cm}
\section*{Acknowledgements}
We thank Aaron Ludlow, Wenting Wang and Jie Wang for helpful discussions
and suggestions. We also thank the anonymous referee for comments
that have helped us improve the paper. MC, CSF
and MS were supported in part by ERC Advanced Investigator grant
COSMIWAY (grant number GA 267291), the Science and Technology
Facilities Council (grant number ST/F001166/1, ST/I00162X/1). LG
acknowledges support from the NSFC grant
(Nos 11133003, 11425312), the Strategic Priority Research Program
'The Emergence of Cosmological Structure' of the Chinese Academy of
Sciences (No. XDB09000000), and a Newton Advanced Fellowship, as well
as the hospitality of the Institute for Computational Cosmology at
Durham University. RAC is a Royal Society University Research Fellow.
JS acknowledges the ERC Grant agreement 278594-GasAroundGalaxies. TT
acknowledges the Interuniversity Attraction Poles Programme initiated
by the Belgian Science Policy Office ([AP P7/08 CHARM]). This work
used the DiRAC Data Centric system at Durham University, operated by
ICC on behalf of the STFC DiRAC HPC Facility (www.dirac.ac.uk). This
equipment was funded by BIS National E-infrastructure capital grant
ST/K00042X/1, STFC capital grant ST/H008519/1, and STFC DiRAC
Operations grant ST/K003267/1 and Durham University. DiRAC is part of
the National E-Infrastructure. We acknowledge PRACE for awarding us
access to the Curie machine based in France at TGCC, CEA,
Bruy\`eres-le-Ch\^atel.

\vspace{-0.2cm}
\bibliographystyle{mnras}
\bibliography{bibliography}

\vspace{-0.4cm}
\appendix
\section{Sample characteristics}
\label{sect:appendix_sample}

\begin{figure}
  \plotone{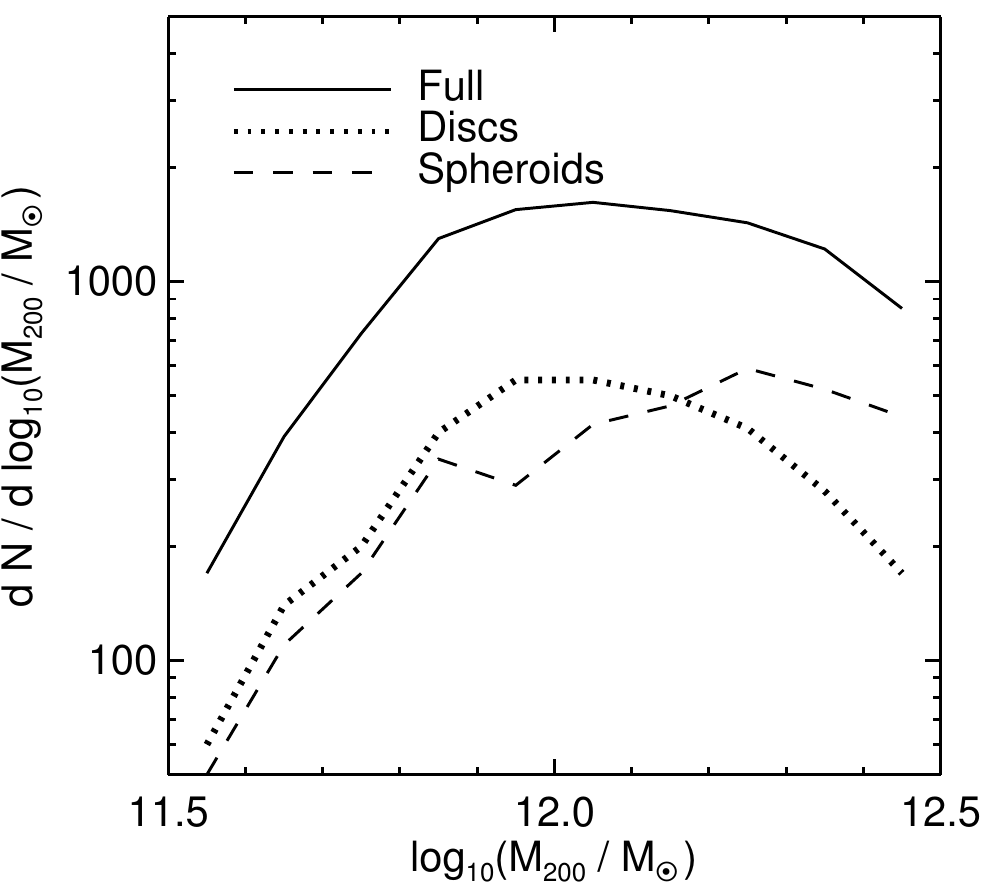}
  \vspace{-.1cm}
  \caption{The number of haloes, $N$, as a function of halo mass for
    the sample of systems that met our selection criteria. Each of the
    disc and spheroid galaxy subsamples contains roughly a third of the
    full sample.
  }
  \label{fig:halo_mass}
  \vspace{-0.4cm}
\end{figure}

\begin{figure}
  \plotone{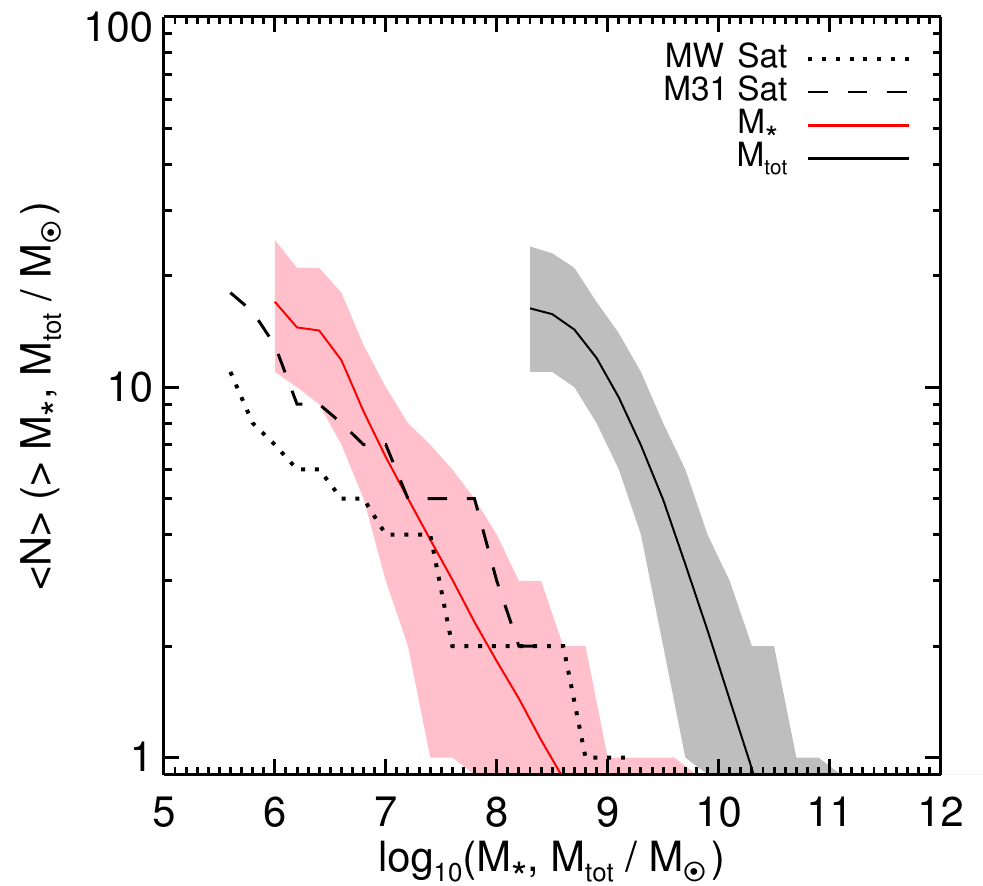}
  \vspace{-.1cm}
  \caption{The mean number of luminous satellites per halo as a
    function of their stellar, $M_\star$, and total, $M_{\rm tot}$,
    masses. The shaded regions indicate the 10th and 90th percentiles
    scatter, while the solid lines indicate the mean value. The dotted and
    dashed lines show the observed stellar mass function within a distance
    of 300 kpc from the MW and the M31.
  }
  \label{fig:sat_mass}
  \vspace{-0.25cm}
\end{figure}

Here, we characterize our sample of MW-like systems in terms of its
distribution of halo masses and its satellite mass function.

Our sample is composed of haloes in the mass range, $M_{200} \in [0.3,
3] \times 10^{12}\Msun$, that contain at least 11 luminous satellite
galaxies within a radius of $300\kpc$. Fig.~\ref{fig:halo_mass} shows
the resulting halo mass distribution for the full sample as well as
for the subsamples split according to the morphology of the central
galaxy. The decrease of the mass distribution below $10^{12}\Msun$ is
due to many low mass haloes not having the required 11 luminous
satellites. The weak decrease at higher masses is due to the
decreasing halo mass function. We also note that while the spheroidal
galaxies have slightly higher halo masses than the discs, we have
checked that this is not the cause behind the difference in alignment
strength of the two populations.

Fig.~\ref{fig:sat_mass} shows the average stellar and total satellite
galaxy mass functions of our sample. Luminous satellites consist of
haloes and subhaloes with at least one star particle, so they can have
stellar masses as low as $\sim 2 \times 10^{6}\Msun$, which
corresponds to the resolution limit of the \eagle{} simulation. The
same satellites have a typical total mass, $M_{\rm tot} \sim 1 \times
10^{9}\Msun$, which, since they are DM dominated, corresponds to $\sim
100$ DM particles. For comparison, we also show the stellar mass
function within a 3D distance of $300\kpc$ from MW and M31, which we
take from the \citet{McConnachie2012} compilation.  We only show the
MW and M31 satellites brighter than $-8.8$ in absolute \textit{V}-band
magnitude since these were the ones used in our study. Considering
fainter satellites would change the observed mass functions only below
a stellar mass of $5\times10^{5}\Msun$. While the \eagle{} satellite
mass function agrees with observations at high masses, it is
systematically higher in the range $M_\star \lesssim
5\times10^{6}\Msun$, especially when compared to the MW. This is an
outcome of selecting only haloes with 11 or more satellites, which
biases our results towards a high satellite count.

\label{lastpage}

\end{document}